\documentclass[conference,onecolumn]{IEEEtran}
\IEEEoverridecommandlockouts
\usepackage{cite}
\usepackage{amsmath,amssymb,amsfonts}
\usepackage{graphicx}
\usepackage{textcomp}
\usepackage{xcolor}
\usepackage{booktabs}
\usepackage{comment}
\usepackage{url}

\graphicspath{{figs/}}

\def\BibTeX{{\rm B\kern-.05em{\sc i\kern-.025em b}\kern-.08em
    T\kern-.1667em\lower.7ex\hbox{E}\kern-.125emX}}

\begin{document}

\title{Smoothing the Ramp, Not the Peak: Scheduling-Induced Power Dynamics of
LLM Inference and Their Grid-Scale Consequences}

\author{
\IEEEauthorblockN{Pan Li\textsuperscript{}}
 \IEEEauthorblockA{
\textit{Tongji University}\\
Shanghai, China \\
pli@tongji.edu.cn}
\and
	\IEEEauthorblockN{Yize Chen\textsuperscript{}}
 \IEEEauthorblockA{
	\textit{University of Alberta}\\
	Edmonton, Alberta, Canada \\
	yize.chen@ualberta.ca}
\and 
	\IEEEauthorblockN{Xia Miao\textsuperscript{}}
\IEEEauthorblockA{\textit{Fova Technology (Suzhou) Co., Ltd} \\
	Suzhou, Jiangsu, China \\
	miaoxia@fovacloud.com}
\and 
	\IEEEauthorblockN{Dai Wang\textsuperscript{}}
\IEEEauthorblockA{\textit{EcoFlare Co., Ltd} \\
	Wuxi, Jiangsu, China \\
	dai@ecoflare.com}

}

\maketitle

\begin{abstract}
Large language model (LLM) inference serving is a fast-growing electricity load whose power
\emph{dynamics} remain uncharacterized from a grid-planning perspective. Using real, measured
GPU power traces, we show that \emph{chunked prefill scheduling}, a latency-motivated
technique already deployed by default in production LLM serving, is a controllable knob that
regulates power ramp rate without touching peak power. Contrary to the intuitive hypothesis
that splitting a long prompt's computation into smaller steps should flatten its power spike,
peak power stays relatively the same while mean ramp rate falls substantially. Critically, this ramp-rate benefit is not a fixed property of the
policy: it grows monotonically with system saturation, and we confirm this along two
independent axes: concurrency (7.0\% at light load to 34.6\% at heavy load, mean-ramp
reduction) and long-prompt (``whale'') request load (from near zero at low whale
incidence to 39.6\% at high whale fraction/size). We translate this single-GPU mechanism into
an operational grid quantity, regulation-reserve procurement, posed and solved as a
chance-constrained problem using a model-free bootstrap directly resampling real measured
power traces. For independently operating servers, chunked prefill reduces the fast-ramping
reserve requirement by an estimated 24\% at heavy load (95\% CIs within 13--34\%) and 12\% at
moderate load, consistently across reliability levels from 95\% to 99.9\%. Repeated with 100 trials per policy on two-GPU tensor-parallel deployments, the reduction is about 36\% for a dense 14B model and 10\% for a sparse mixture-of-experts model. The benefit
holds for ramps resolved at up to 2\,s and persists at a reduced 16--20\% when servers' load
changes are synchronized. Together, these results identify a no-cost lever for fast-timescale
demand shaping that is available today, whose benefit is largest when data centers run hottest.
\end{abstract}


\section{Introduction}
\label{sec:intro}

Global data center electricity demand is projected to more than double by 2030, with AI the
primary driver \cite{iea2025energyai,chen2026gridimpacts}. Most attention to this growth
addresses \emph{training} workloads, where synchronized collective communication across large
GPU clusters produces well-documented, megawatt-scale power oscillations
\cite{choukse2025powerstabilization,llama3herd2024}. \emph{Inference serving}, the workload
that runs continuously once a model is deployed, has received comparatively little grid-side
characterization, despite a structurally different demand pattern: request-level heterogeneity that results from a mix of short and long prompts rather than fleet-wide synchronized compute phases. This
heterogeneity is a direct consequence of how production LLM applications construct prompts. For example, retrieval-augmented generation, chat history, and pasted documents occasionally add a long
prefill to an otherwise short-request stream. It is precisely what motivated chunked-prefill
scheduling \cite{agrawal2023sarathi,agrawal2024sarathiserve}, making it the natural workload
for studying that policy's other consequences.

This paper asks a narrower, concrete question: does chunked prefill, used by default in production serving systems such as vLLM
\cite{agrawal2024sarathiserve} specifically to bound tail latency, have a measurable effect
on a serving GPU's power \emph{shape}? From our experiments and rigorous analysis, the answer is yes, but not in the way latency-motivated
intuition predicts. The natural hypothesis, that slicing a long prompt's computation into many
small scheduler steps should flatten its power spike, turns out to be \emph{wrong}. We find that peak
power is essentially unmoved, while ramp rate and duty cycle shift substantially. A policy
adopted for an entirely unrelated reason protecting short-request tail latency turns out
to double as a ramp-rate demand-shaping lever, at essentially no new deployment cost. The
remainder of this paper generalizes this finding to fleet scale and applies it to an actual
grid-operations decision, summarized in the contributions below.
\begin{enumerate}
\item A controlled measurement showing chunked prefill scheduling lowers mean ramp rate
substantially without moving peak power, localized directly to
long-prompt (``whale'') processing windows.
\item A ramp-rate reduction that grows monotonically with system saturation, confirmed
independently along two axes: concurrency (7.0\% to 34.6\% mean-ramp reduction,
light-to-heavy load) and whale-request load (a whale-fraction $\times$ whale-size grid showing
the same monotonic growth), establishing a genuine, saturation-dependent property of
the scheduling policy.
\item A direct translation into an operational grid quantity by solving
regulation-reserve procurement as a chance-constrained problem, using a model-free, nested
bootstrap over 100 real measured traces per policy: chunked prefill needs about 24\% less
fast-ramping reserve than mono at heavy load and 12\% at moderate load, with confidence
intervals, a closed-form explanation, and sensitivity to cross-server correlation and ramp
timescale (Section~\ref{sec:cf-reserve}).
\item A replication of the full reserve pipeline on a 14B dense model and an MoE model under
tensor parallelism, again with 100 trials per policy (Section~\ref{sec:reserve-tp2}).
\end{enumerate}

\begin{figure*}[t]
\centering
\includegraphics[width=\textwidth]{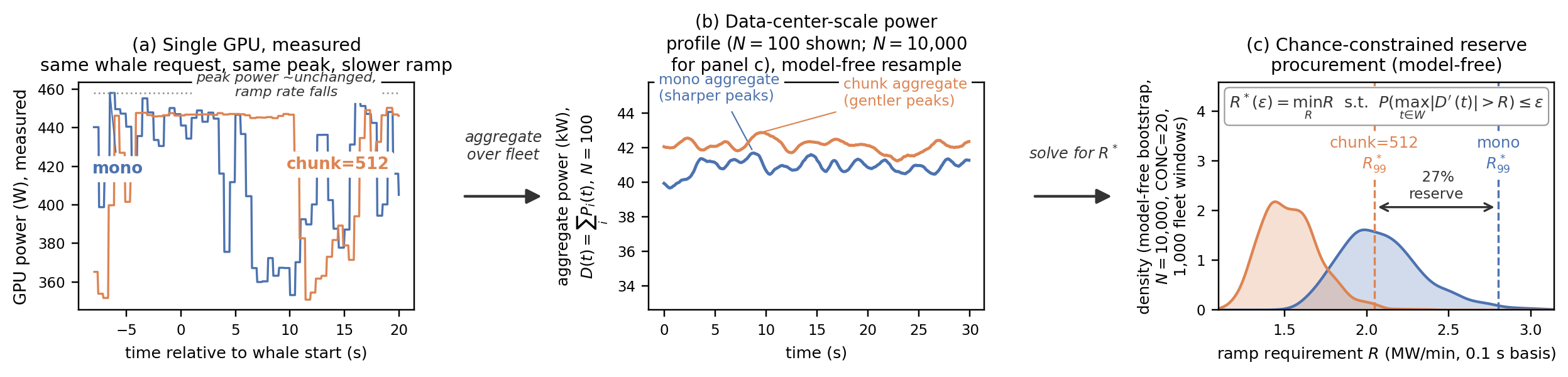}
\caption{Overview of the paper's argument from single GPU to grid-operations decision (illustrative). \emph{(a)} Chunked prefill cuts ramp rates substantially (Section~\ref{sec:ramp}).
\emph{(b)} Per-GPU effect propagates to fleet scale via bootstrapping.
\emph{(c)} At heavy load, chunked prefill requires about 24\% less fast-ramping reserve than
mono (Table~\ref{tab:reserve}; the panel shows one realization of 1,000 fleet windows).}
\label{fig:overview}
\end{figure*}

The paper is organized as the following. Section~\ref{sec:related} covers background and related work. Section~\ref{sec:problem} states
the grid-operations decision problem that motivates everything
that follows. Section~\ref{sec:setup} details the hardware setup and workload and reports
the single-GPU power-shape finding. Section~\ref{sec:load-dependence} shows this benefit grows
monotonically with system saturation along two independent axes, and
Section~\ref{sec:cf-reserve} solves the problem posed in
Section~\ref{sec:problem} using a model-free bootstrap over the measured traces.
Section~\ref{sec:conclusion} concludes with further works. Figure~\ref{fig:overview} gives a high-level overview of the paper's argument.

\section{Background and Related Work}
\label{sec:related}

\subsection{LLM serving scheduling} 
\begin{figure}[t]
\centering
\includegraphics[width=\columnwidth]{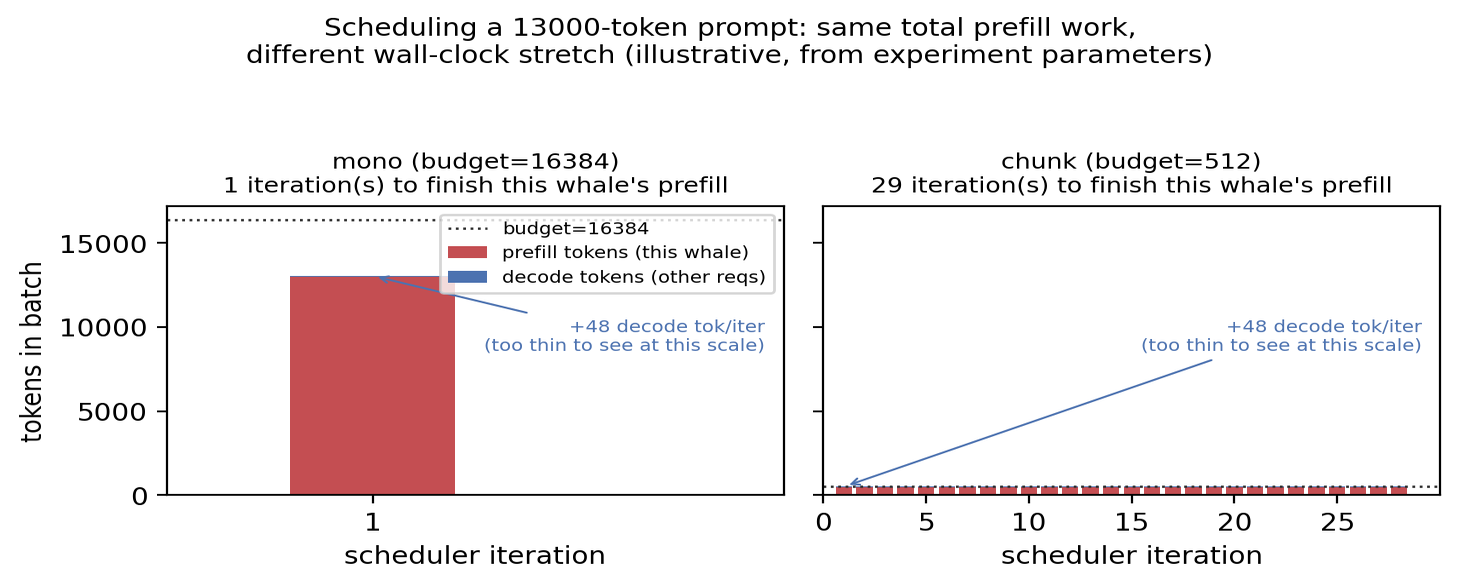}
\caption{Illustrative schematics of chunked-prefill batch composition.
Across a representative 13,000-token whale's full
prefill, mono finishes in one large iteration while chunk=512 needs 29 small iterations -- the
same total prefill work stretched over far more scheduler rounds. The decode segment (48
tokens/iteration, from concurrent short requests) is genuinely stacked atop the prefill bar in
both panels, annotated rather than exaggerated: at $\le$0.3\% of the shared 16,384-token axis,
it is sub-pixel and would otherwise be invisible.}
\label{fig:prefill-decode-schematic}
\end{figure}

Serving one LLM request has two phases with different compute
profiles: \emph{prefill}, a single, highly parallel pass processing the entire input prompt at
once (compute-intensive, proportional to prompt length), and \emph{decode}, the subsequent
step-by-step generation of output tokens one at a time (cheap per step, but many steps for a
long response). A GPU serving many concurrent requests batches decode steps from in-progress
requests together. Chunked prefill addresses what happens when a new, long prompt's prefill
needs to run: if left unchunked, it can dominate an entire scheduler iteration (one discrete
round of GPU compute) and stall every other request's decode step behind it. Chunked prefill
\cite{agrawal2023sarathi,agrawal2024sarathiserve} splits a long prompt's prefill across
multiple scheduler iterations, interleaved with ongoing decode work, specifically to bound the
worst-case inter-token latency any concurrent request experiences, and is now default in
mainstream serving frameworks. Figure~\ref{fig:prefill-decode-schematic} illustrates the
mechanism driving our power-shape finding (Section~\ref{sec:ramp}): capping the per-iteration
prefill budget caps how much of any single scheduler step a whale prompt can dominate, leaving
more of that step's batch available to concurrent decode work, at the cost of needing many
more steps to finish.

\subsection{Related work} 
Training-side power behavior is comparatively mature, with synchronized
collective-communication phases producing documented, grid-relevant power oscillations
\cite{choukse2025powerstabilization,ko2025wideareaoscillations,llama3herd2024}; yet the literature
does not address inference serving, where the load driver is per-request heterogeneity rather
than synchronized compute phases. Inference-side characterization is newer
\cite{wilkins2026serverstosites,majumder2026workloadcomposition}, and closest to
\cite{liang2026inferenceflexibility}, which also targets ramp rate but as \emph{exogenous
flexibility offsetting a co-located training job's ramp}, not a characterization of inference's
\emph{own} scheduling-induced power shape. We differ by treating scheduling policy itself as
the object of study, and by grounding fleet-scale extrapolation in a
diversity/coincidence-factor formalism, a distribution-planning tool most visibly applied
recently to EV-charging demand \cite{bollerslev2022coincidence}, which we are, to our
knowledge, the first to apply to an LLM-serving load calibrated against a real measured trace. Separately, our finding that peak
power resists chunking's influence while duty cycle shifts is consistent with
evidence that server and GPU power is far from proportional to activity, staying high
even when little work is visible \cite{barroso2007energyproportional,lei2026executionidle} where
chunking changes \emph{when} the GPU is saturated, not \emph{whether} it can be
power-proportional at all.

\section{Problem Formulation}
\label{sec:problem}

Utilities routinely procure fast-responding ramp-following reserve/capacity.  For example, generation or storage capacity
can be held specifically to track a load's short-timescale fluctuations, and sized against how
\emph{fast} a load can ramp, not only how large it can grow \cite{ferc2024ancillary}. Slower
facility-level ramping is understood to reduce how much of this reserve a utility must hold
against a given load. This paper's central operational question is exactly how much less, for a
real, measured LLM-serving load under an already-deployed scheduling policy, posed here as an
explicit capacity-scheduling problem and solved directly from a fleet scale ramp distribution,
rather than only compared against a single interconnection limit.

\textbf{Problem statement.} Let $R$ be the \emph{ramp-following requirement} (MW/min), the rate
at which a utility's fast-response reserve must be able to move to track the load, distinct
from the total capacity (MW) that reserve must hold, which this paper does not address. We
express $R$ in MW/min but measure $D'(t)$ over short (0.1--1\,s) differences, so $R$ is a
\emph{fast}-ramp requirement; Section~\ref{sec:reserve-scope} examines longer timescales. We scope
the problem to a single facility's demand against its own point-of-interconnection agreement, e.g.\ the $10$\,MW/min limit already imposed by the Alberta Electric System Operator on
transmission-connected AI data centers \cite{liang2026inferenceflexibility}, not to utility-wide
balancing-area reserve, since the facility level is where a load's own scheduling policy has
direct leverage over $D(t)$. A system operator must ensure its reserve fleet can ramp fast enough
that the load's realized ramp exceeds that capability only rarely, i.e.\ it must solve

\begin{equation}
\label{eq:reserve-opt}
R^*(\epsilon) \;=\; \min_R \; R \quad \text{s.t.} \quad P\Big(\max_{t\in W}|D'(t)| > R\Big) \le \epsilon,
\end{equation}

\noindent for a target reliability level $1{-}\epsilon$ (e.g.\ 95\%, 99\%, 99.9\%) over an
operating window $W$ (30\,s here). The optimum
$R^*(\epsilon)$ is exactly the $(1{-}\epsilon)$-quantile of the per-window peak-ramp distribution of the
aggregate fleet-scale demand $D(t)$, which is a distribution derived directly from
measured single-GPU traces (Section~\ref{sec:ramp}) propagated to fleet scale via a model-free
bootstrap (Section~\ref{sec:cf-reserve}). Section~\ref{sec:cf-reserve} solves
Eq.~\eqref{eq:reserve-opt} numerically once that distribution is in hand.

\section{Experimental Setup}
\label{sec:setup}
\subsection{Hardware and LLM setup}
We experiment with Qwen2.5-Coder-7B-Instruct served via vLLM on a single NVIDIA GeForce RTX 4090
(24\,GB, 450\,W, no tensor parallelism). Section~\ref{sec:reserve-tp2} and
Appendix~\ref{app:tp2-14b} follow up on a larger dense model and an MoE model, both split across
two GPUs via tensor parallelism. We also want to note that this is a consumer-grade accelerator
rather than a datacenter-class part (e.g., A100, H100). However, we expect the qualitative smoothing
mechanism to carry over, since it follows from how the scheduler slices compute rather than from
board-specific power behavior, but the ramp rate and duty cycle magnitudes in the scope of this paper should be
read as specific to this hardware class.

We conducted three scheduling configurations
which differ only in the
step-wide prefill token budget: 16384 (effectively unchunked, ``mono''), 2048, and 512. The
workload is drawn from ShareGPT\cite{sharegpt} conversations, with each trial sampling a
distinct, non-overlapping slice of 200 requests. Requests
follow a bimodal length mixture: a short-prompt majority and a long-prompt (``whale'')
minority, with prefix padding using unique, uncacheable content. This models a mix of general
chat and agentic/tool-augmented traffic grounded to \cite{wang2025burstgpt, zhu2026tracelab}.

Section~\ref{sec:ramp} reports the mechanism at one
heavily-loaded reference condition ($\sim$20 requests continuously in flight, chosen to mimic
production serving at a representative, cost-motivated utilization level which is the same condition
used in Section~\ref{sec:load-dependence}), drawn from
a bimodal length mixture with an 85\% short-prompt population and a 15\% long-prompt population
in an 18,000--50,000 character range ($\approx$5.5k--15.5k tokens). This mixture
exaggerates a skew already present in real ShareGPT-derived traffic (retrieval-augmented
context, chat history, pasted documents), with the same request sequence replayed against every
configuration within a trial. Other traffic patterns (Poisson arrivals, Pareto-tailed
whale sizes, a whale-free condition) are tested in Appendix~\ref{app:robustness}.

\subsection{Instrumentation}
Power is sampled via an NVML side-car at 50\,ms resolution, cross-checked against chassis-level
BMC sensors, with GPU clock locked to remove DVFS-driven variance; each trial discards an initial
warmup burst. The NVML reading itself updates only every
$\approx$0.5\,s (median) on this GPU and may be averaged by the driver over up to 1\,s, so the
traces are already low-pass filtered at the sensor; we resample them to a 0.1\,s grid. Further robustness checks reproduce the
same qualitative pattern and establish the boundaries of the effect. For interested readers please refer to Appendix~\ref{app:robustness}.

\subsection{Single chip observation: Energy Conservation and Power-Ramp Reshaping}
\label{sec:ramp}

\textbf{Energy is approximately conserved.}\label{sec:energy} Scheduling policy should not
change the total compute performed, only its temporal arrangement, so energy per output token
should be near-invariant across configurations. This holds closely at the moderate chunk size
(701.7 kJ/1k output tokens at 2048 vs.\ 701.7 at mono, +0.0\%), with a modest cost at the most
aggressive setting (715.5 at 512, +2.0\%), consistent with per-iteration scheduling overhead
scaling with the number of rounds a long prompt is split across.

\textbf{Peak power does not flatten, but the temporal shape does change.} Peak power during
long-prompt processing saturates near the GPU's ceiling in all three configurations (463.9\,W
mono, 470.8\,W at 2048, 470.9\,W at 512 with differences $\le$1.5\%): the GPU commits the same
total compute to a long prompt regardless of how it is sliced in time. What does change, systematically and
reproducibly, is the mean instantaneous ramp rate ($|dP/dt|$) and the near-ceiling duty cycle
as shown in Table~\ref{tab:ramp_avg}, reported as mean $\pm$ std across 10 trials.

\begin{table*}[htbp]
	\caption{Ramp rate and near-ceiling duty cycle by scheduling budget, mean $\pm$ std across 10 trials.}
	\label{tab:ramp_avg}
	\centering
	\small
	\begin{tabular}{@{}lcccccc@{}}
		\toprule
		 mono ramp (W/s) & 2048 ramp (W/s) & 512 ramp (W/s) & mono$\ge$thresh & 2048$\ge$thresh & 512$\ge$thresh \\
		\midrule
		 45.6$\pm$2.6 & 43.1$\pm$3.0 & 29.8$\pm$3.2 & 43.7\%$\pm$3.7 & 48.9\%$\pm$4.1 & 60.0\%$\pm$4.5 \\
		\bottomrule
	\end{tabular}
\end{table*}

\begin{figure*}[t]
	\centering
	\includegraphics[width=0.8\textwidth]{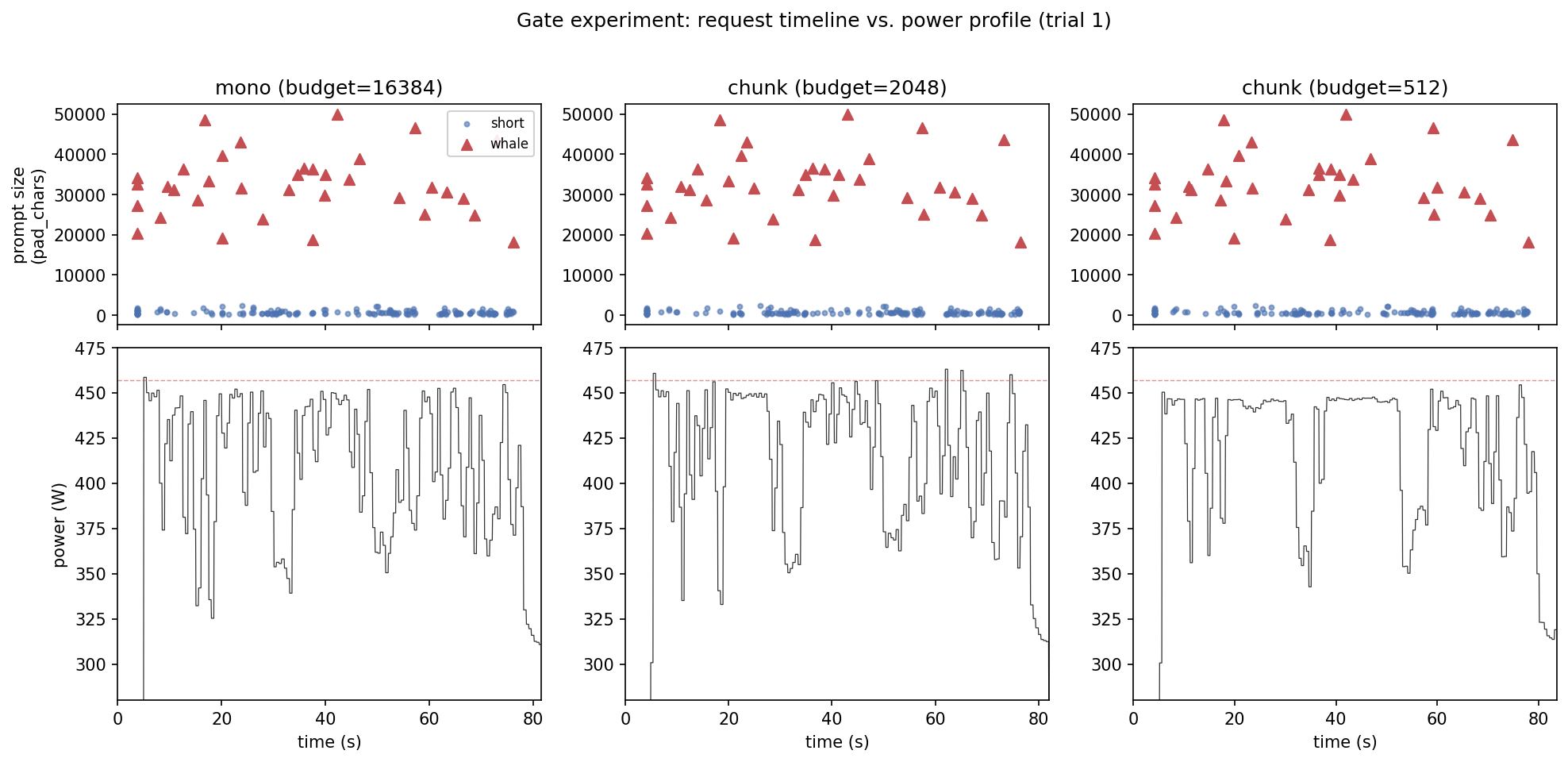}
	\caption{Request arrival/prefill-size timeline (top) against the GPU power trace (bottom) for
		the same seeded request sequence under mono/16384, chunk/2048, and chunk/512 (trial 1,
		concurrency 20, widened request-size distribution). Red triangles mark whale requests,
		spanning $\sim$18k--50k characters. The mono panel shows sharp, frequent drops toward
		$\sim$300\,W across the trace; the 512 panel shows a sustained, smoother plateau with fewer,
		shallower excursions -- the direct visual counterpart of the ramp-rate result in
		Table~\ref{tab:ramp_avg}.}
	\label{fig:timeline}
\end{figure*}

As chunking gets more aggressive, mean ramp rate falls (mono$\to$512: $\sim$46$\to$30\,W/s,
$\sim$35\%) and the number of elevated-power episodes falls (fewer, longer excursions), while
the \emph{fraction of time} spent near the ceiling rises ($\sim$44\% at mono to $\sim$60\% at
chunk=512). This holds specifically within long-prompt processing windows (near-ceiling time is
essentially zero, under 0.1\%, outside them for every configuration). Figure~\ref{fig:timeline} makes this visually concrete: the mono trace
shows sharp, frequent drops toward baseline, while chunk=512 sustains a smoother, longer
near-ceiling plateau with far fewer deep excursions. 

\textbf{Mechanism.} Chunking prefill requests spreads that same total work over more wall-clock time,
interleaved with concurrent short-request decode work. Because this smoothing works by
interleaving a whale's own chunks with \emph{other} concurrently in-flight requests, it needs
concurrent demand to interleave with: the effect is markedly more pronounced the more saturated
the server is (Section~\ref{sec:load-dependence}).

\textbf{Why this matters, and what it costs.} Ramp rate, not peak magnitude, is the quantity
many interconnection standards regulate for large flexible loads (battery storage, EV
fast-charging) \cite{liang2026inferenceflexibility}. A $\sim$35\% reduction, from a policy protecting short-request tail latency, is a demand-shaping lever at essentially no additional
cost or new infrastructure. 
Short-request tail latency falls in
lockstep with ramp rate as chunking gets more aggressive, at a cost to
the whale request's own time-to-first-token (Appendix~\ref{app:latency}), so the ramp-rate
benefit rides along with a latency benefit the policy was already paying for.

\section{Load-Dependence Across Concurrency and Whale Load}
\label{sec:load-dependence}

Section~\ref{sec:ramp} establishes the ramp-reduction mechanism at one meaningfully-loaded
reference condition. Here we ask whether the benefit is a fixed property of chunked scheduling
or depends on how saturated the server is, along two independent axes: closed-loop concurrency
and whale-request load. Both use single-server mean/p99 ramp-rate reduction (mono
$\to$ chunk=512) as the metric, computed identically to Section~\ref{sec:ramp}. 

\textbf{Concurrency.} Table~\ref{tab:load-curve} sweeps concurrency from light (6)
through moderate (10) to heavy (20) load, at the default whale mixture (15\% whale fraction,
18k--50k character / $\sim$5.5k--15.5k token size). Mean-ramp reduction grows nearly five-fold across this range, and
the p99 tail statistic follows the same monotonic trend.

\begin{table}[htbp]
\caption{Single-server ramp reduction (mono $\to$ chunk=512) vs.\ closed-loop concurrency, 10
trials/arm each, diverse conversations}
\label{tab:load-curve}
\centering
\small
\begin{tabular}{@{}lcc@{}}
\toprule
Concurrency & Mean-ramp & p99-ramp \\
\midrule
6 (light)     & 7.0\%  & 7.4\%  \\
10 (moderate) & 17.0\% & 11.6\% \\
20 (heavy)    & 34.6\% & 18.6\% \\
\bottomrule
\end{tabular}
\end{table}

\textbf{Whale fraction and size.} Table~\ref{tab:whale-grid} sweeps whale fraction
(5\%/15\%/30\%) and whale size ($\sim$8k/$\sim$14k tokens) at a fixed, moderate concurrency
(10) on the narrow (near-constant-per-cell) whale-size distribution, holding concurrency and
every other harness setting identical to the sweep above (only whale-size \emph{variance}
is deliberately narrower to isolate the fraction/size axes cleanly). The same monotonic
pattern holds along this orthogonal axis: at the smaller whale size, reduction is
near zero at low whale fraction (5\%: $-$3.2\%) and grows to 19.1\% at high fraction (30\%); at
the larger whale size, the same sweep runs from 2.7\% to 39.6\%.

\begin{table}[htbp]
\caption{Single-server ramp reduction vs.\ whale fraction $\times$ whale size, concurrency 10,
10 trials/arm each, narrow per-cell whale-size distribution.}
\label{tab:whale-grid}
\centering
\small
\begin{tabular}{@{}lcc@{}}
\toprule
Whale frac.\ / size & Mean-ramp & p99-ramp \\
\midrule
5\% / $\sim$8k    & $-$3.2\% & 5.8\%  \\
5\% / $\sim$14k   & 2.7\%  & 0.1\%  \\
15\% / $\sim$8k   & 8.8\%  & 9.0\%  \\
15\% / $\sim$14k  & 19.0\% & 8.5\%  \\
30\% / $\sim$8k   & 19.1\% & 12.1\% \\
30\% / $\sim$14k  & 39.6\% & 15.9\% \\
\bottomrule
\end{tabular}
\end{table}

\textbf{Mechanism.} Both axes are consistent with Section~\ref{sec:ramp}'s interleaving
mechanism: chunking's smoothing effect works by interleaving a whale's own chunks with
\emph{other} concurrently in-flight requests: more
concurrency, or more/larger whale requests generating more overlapping prefill demand leads to material effect.

\section{Application: Ramping-Reserve Procurement}
\label{sec:cf-reserve}

Section~\ref{sec:problem} posed the ramping-reserve procurement problem
(Eq.~\eqref{eq:reserve-opt}). We solve it by model-free extrapolation from real, measured
single-server power traces, rather than from a measured multi-server fleet or from the parametric
coincidence-factor model of Appendix~\ref{app:ou-model}, whose Gaussian tail runs measurably
lighter than real hardware's ($\approx6\times$ at the 99th percentile).

\subsection{Method and Uncertainty}
\label{sec:reserve-method}

\textbf{Fleet construction.} We build $N{=}10{,}000$ virtual servers, each drawing an
independent, randomly phase-shifted 30\,s window from a library of 100 real trials per policy,
and sum them to $D(t)=\sum_i P_i(t)$. Trials were collected in alternating 10-trial batches across
the two scheduling policies, so that slow drift over the multi-hour collection session cannot be
confounded with the policy comparison. Each fleet window yields one sample of the peak ramp
$\max_{t\in W}|D(t{+}\Delta)-D(t)|/\Delta$, with $\Delta=0.1$\,s unless stated otherwise.

\textbf{Start-up transient (correction to v1).} We discard the first 8\,s of every trace after
the 300\,W trim. Each trial begins with a $\approx$5\,s start-up ramp ($\approx$300$\to$410\,W
on the single GPU) which, because all virtual servers draw their phases from the same range,
adds \emph{coherently} across the fleet. With uniform phases the expected fleet step at window
offset $j$ is $\frac{N}{K}\big(\bar m(j{+}K)-\bar m(j)\big)$, where $\bar m$ is the library-mean
trace and $K$ the number of admissible phases; for the concurrency-20 library this deterministic
ramp reaches 2.9\,MW/min without the skip and at most 0.5\,MW/min with it, against $R^*$ of
2--4\,MW/min. Version~1 of this preprint retained the transient, which added a similar deterministic
ramp to both policies' $R^*$, diluting the relative reduction and widening its uncertainty.

\textbf{Uncertainty.} With 100 traces underlying 10{,}000 servers, the dominant uncertainty is
which trials populate the library. We therefore use a nested bootstrap: each of $B{=}100$ outer
replicates resamples both libraries with replacement, and an inner Monte Carlo of 1{,}000 fleet
windows per policy gives $R^*(\epsilon)$ as an \emph{empirical} quantile. We report the mean
reduction $1-R^*_{\text{chunk}}/R^*_{\text{mono}}$ and its 2.5--97.5 percentile interval across
replicates. No parametric tail model is needed up to 99.9\%. As a check, a Gumbel fit to 30
window maxima (the approach of v1) overstates the empirical 99.9\% quantile of 4{,}000 windows by
0.3--9.2\% across the four policy/load combinations, i.e.\ conservatively.

\subsection{Results}

\begin{table}[htbp]
\caption{Fast-ramping reserve reduction (mono $\to$ chunk=512), $N{=}10{,}000$, 0.1\,s ramp,
independent servers, Qwen2.5-Coder-7B on one GPU: mean [95\% CI] over 100 nested-bootstrap
replicates; mono $R^*$ in MW/min.}
\label{tab:reserve}
\centering
\footnotesize
\setlength{\tabcolsep}{4pt}
\begin{tabular}{@{}lcccc@{}}
\toprule
 & \multicolumn{2}{c}{Concurrency 20 (heavy)} & \multicolumn{2}{c}{Concurrency 10 (moderate)} \\
\cmidrule(lr){2-3}\cmidrule(lr){4-5}
Reliability & Reduction & $R^*_{\text{mono}}$ & Reduction & $R^*_{\text{mono}}$ \\
\midrule
95\%   & 24.3\% [17.8, 28.5] & 2.62 & 11.6\% [8.9, 14.4] & 3.22 \\
99\%   & 24.1\% [16.1, 29.4] & 2.88 & 11.7\% [7.5, 16.1] & 3.55 \\
99.9\% & 23.6\% [13.2, 33.5] & 3.17 & 11.8\% [3.6, 21.1] & 3.91 \\
\bottomrule
\end{tabular}
\end{table}

At heavy load, chunk=512 needs about 24\% less fast-ramping reserve than mono at every
reliability level, with every interval excluding zero (Table~\ref{tab:reserve}); at moderate
load the reduction is about 12\%, mirroring the load dependence of
Section~\ref{sec:load-dependence}. A second, independently collected heavy-load library with
twice as many requests per trial (Section~\ref{sec:reserve-scope}) gives 27.0\% [24.0, 30.3],
26.9\% [22.6, 31.0] and 26.4\% [16.8, 32.9] at the three levels, consistent with
Table~\ref{tab:reserve}. This reduction is not free: at chunk=512, whale
time-to-first-token rises over mono even as short-request tail latency improves
(Appendix~\ref{app:latency}), so the grid-side benefit and this latency cost should be weighed
together.

\textbf{Why the reduction is stable across $N$ and $\epsilon$.} For independent servers, $D'(t)$
is a sum of $N$ independent per-server ramps, approximately Gaussian with standard deviation
$\sqrt{N}\sigma$, where $\sigma$ is the single-server RMS ramp. Hence
$R^*\approx z_\epsilon\sqrt{N}\sigma$, with a quantile factor $z_\epsilon$ (for the maximum over a
window) approximately common to both policies, and the reduction
$\approx1-\sigma_{\text{chunk}}/\sigma_{\text{mono}}$ is independent of $N$ and $\epsilon$. From
the per-trace RMS ramps this predicts 26.3\% (heavy) and 11.4\% (moderate), matching the
bootstrap. The fleet-level reduction is smaller than the $\approx$35\% mean-ramp reduction of
Section~\ref{sec:ramp} because the RMS is dominated by the largest single-server steps, which
chunking reduces less (p99 column of Table~\ref{tab:load-curve}).

\subsection{Larger and Sparse Models Under Tensor Parallelism}
\label{sec:reserve-tp2}

Version~1 reported only 3-trial pilots for a larger model and an MoE model, evaluated with the
parametric model (Appendix~\ref{app:tp2-14b}). We now repeat the full reserve pipeline on both:
Qwen2.5-Coder-14B-Instruct (dense) and Qwen1.5-MoE-A2.7B-Chat (14.3B total, 2.7B active
parameters), each split across two RTX 4090s with tensor parallelism (TP=2, PCIe only), at
concurrency 20 with the same widened whale mixture as Table~\ref{tab:reserve}. Each model has
100 trials per policy, collected in alternating 10-trial batches on two GPU pairs, each pair
contributing equally to both policies; each trial replays its own slice of 200 conversations, the
same slice under both policies; power is the sum over the two GPUs of a pair. No request
was preempted in any trial. The fleet model is unchanged, with each virtual server now a
two-GPU TP group.

\begin{table}[htbp]
\caption{Reserve reduction (mono $\to$ chunk=512) for TP=2 models at concurrency 20,
$N{=}10{,}000$ TP groups, 0.1\,s ramp, independent servers: mean [95\% CI] over 100
nested-bootstrap replicates; mono $R^*$ in MW/min. ``Single-server'': mean-ramp reduction of the
library traces; ``CLT'': $1-\sigma_{\text{chunk}}/\sigma_{\text{mono}}$.}
\label{tab:reserve-tp2}
\centering
\footnotesize
\setlength{\tabcolsep}{4pt}
\begin{tabular}{@{}lcccc@{}}
\toprule
 & \multicolumn{2}{c}{14B dense, TP=2} & \multicolumn{2}{c}{MoE A2.7B, TP=2} \\
\cmidrule(lr){2-3}\cmidrule(lr){4-5}
 & Reduction & $R^*_{\text{mono}}$ & Reduction & $R^*_{\text{mono}}$ \\
\midrule
95\%   & 35.9\% [32.8, 38.3] & 2.56 & 10.4\% [5.0, 15.0] & 2.26 \\
99\%   & 36.0\% [32.2, 39.5] & 2.82 & 10.3\% [4.0, 15.5] & 2.50 \\
99.9\% & 35.8\% [29.5, 41.6] & 3.10 & 10.7\% [$-$0.4, 19.2] & 2.76 \\
\midrule
Single-server & \multicolumn{2}{c}{31.1\%} & \multicolumn{2}{c}{3.5\%} \\
CLT           & \multicolumn{2}{c}{35.8\%} & \multicolumn{2}{c}{11.5\%} \\
\bottomrule
\end{tabular}
\end{table}

For the dense 14B model, chunk=512 needs about 36\% less fast-ramping reserve at every
reliability level, with intervals of 29.5--41.6\% even at 99.9\%
(Table~\ref{tab:reserve-tp2}). This is larger than the 24\% of the 7B single-GPU deployment and
again matches the CLT prediction (35.8\%). Unlike the 7B case, the fleet-level reduction
exceeds the single-server mean-ramp reduction (31.1\%), because under TP=2 chunking shrinks the
largest per-server steps at least as much as typical ones. For the MoE model the reduction is
about 10\% at 95\% and 99\% reliability, with intervals excluding zero. At 99.9\% the empirical
interval reaches zero ($-$0.4\%), so the benefit is not resolved at that level by 100 traces; the
Gumbel check gives 10.4\% [5.9, 15.2]. The MoE single-server mean-ramp reduction is only 3.5\%:
chunking trims the occasional large step that sets the RMS more than it trims typical steps, and
the small active-parameter count leaves less sustained prefill compute to interleave. Peak power
changes little under chunking: the mean per-trial peak falls 1.3\% (14B) and 2.9\% (MoE), and
the maximum over 100 trials changes by 0.0\% and $-$3.6\%. The 8\,s start-up skip matters here as well: without it the coherent start-up
ramp reaches 4.5\,MW/min (14B) and 3.9\,MW/min (MoE) at $N{=}10{,}000$; with it, at most 0.3 and
0.4\,MW/min.

\textbf{Correction to the v1 MoE pilot.} Version~1 reported a 19.8\%$\pm$12.8 mean-ramp reduction
for the MoE model at concurrency 20 from 3 trials (Appendix~\ref{app:tp2-14b}). The same per-trial
metric over the 100-trial library gives 4.4\%$\pm$9.7 (mean $\pm$ std across trials, full
traces). Recomputing the v1 pilot traces with our current code reproduces its value (20.0\%;
trials 39.0/19.9/1.1\%), so the difference reflects sampling, not a change of metric: with a
per-trial spread of about 10 points, 3 trials cannot resolve an effect of this size. At the fleet
level the MoE benefit is real (about 10\%), but the single-server effect is a fraction of the pilot
estimate.

\subsection{Scope of the Fleet Extrapolation}
\label{sec:reserve-scope}

\begin{table}[htbp]
\caption{Sensitivity of the reserve reduction, 7B, concurrency 20, on the second library of
100 trials per policy with 400-request trials ($\approx$111\,s per trace after the start-up skip):
mean [95\% CI] over 100 nested-bootstrap replicates, 1{,}000 (baseline) or 100 fleet windows each. $\Delta$: ramp differencing interval;
$\tau$: first-order low-pass time constant applied to $D(t)$ before 0.1\,s differencing; $s$:
fraction of servers sharing one trace and phase.}
\label{tab:sensitivity}
\centering
\footnotesize
\begin{tabular}{@{}llcc@{}}
\toprule
Factor & Setting & 95\% level & 99\% level \\
\midrule
Baseline      & $\Delta=0.1$\,s & 27.0\% [24.0, 30.3]      & 26.9\% [22.6, 31.0] \\
Ramp interval & $\Delta=0.5$\,s & 26.7\% [20.7, 33.8]      & 26.7\% [16.3, 36.5] \\
($s=0$)       & $\Delta=1$\,s   & 23.5\% [16.3, 29.9]      & 22.7\% [11.7, 31.3] \\
              & $\Delta=2$\,s   & 13.8\% [2.5, 21.9]       & 13.1\% [$-$4.2, 27.2] \\
              & $\Delta=5$\,s   & 13.7\% [0.2, 26.0]       & 13.7\% [$-$4.8, 31.6] \\
              & $\Delta=10$\,s  & 13.8\% [$-$7.7, 31.3]    & 11.8\% [$-$11.6, 28.8] \\
Low-pass      & $\tau=0.5$\,s   & 22.3\% [16.9, 28.6]      & 21.8\% [11.3, 30.6] \\
($s=0$)       & $\tau=1$\,s     & 18.6\% [9.8, 26.9]       & 18.0\% [3.9, 28.0] \\
              & $\tau=2$\,s     & 16.8\% [6.3, 25.4]       & 16.7\% [2.1, 28.7] \\
              & $\tau=5$\,s     & 15.9\% [$-$0.6, 30.0]    & 15.5\% [$-$9.0, 30.7] \\
Correlation       & $s=0.001$ & 27.2\% [20.8, 32.4] & 27.5\% [16.6, 35.0] \\
($\Delta=0.1$\,s) & $s=0.01$  & 19.9\% [12.3, 25.8] & 20.3\% [7.8, 33.3] \\
                  & $s=0.05$  & 17.0\% [11.9, 22.6] & 17.4\% [7.9, 29.6] \\
                  & $s=0.2$   & 16.2\% [10.6, 23.8] & 16.3\% [8.9, 24.1] \\
                  & $s=1$     & 16.0\% [7.8, 23.6]  & 16.2\% [9.2, 26.2] \\
\bottomrule
\end{tabular}
\end{table}

\textbf{Correlated load changes.} Independent, phase-shifted resampling captures the diversity
of fluctuations within steady closed-loop serving, but not events that move many servers
together: traffic surges, diurnal swings, failover and load rebalancing, or coordinated batch-job
starts. Such common-mode changes scale with $N$ rather than $\sqrt{N}$, are absent from our
stationary traces, and are not smoothed by chunking, which acts within a server. We emulate
partial synchronization by giving a fraction $s$ of servers the same trace and phase
(Table~\ref{tab:sensitivity}). Synchronization raises the absolute requirement steeply: mono
$R^*$ at 99\% grows $2.0\times$, $8.0\times$, $32\times$ and $\approx$158$\times$ at $s{=}0.01$,
$0.05$, $0.2$ and $1$. The relative reduction falls to 16--20\% at 95--99\% reliability and
13--17\% at 99.9\%, with every interval excluding zero, because synchronized servers still run
chunked schedules. (On the shorter library of Table~\ref{tab:reserve}, the 99.9\% synchronized
estimate was not resolved.) The independent-server numbers in
Tables~\ref{tab:reserve} and~\ref{tab:reserve-tp2} are thus a best case; a fleet with correlated
load changes needs a much larger reserve, of which chunking removes a smaller but still positive
share. A homogeneous library (one model and one GPU type per table) also understates the
diversity of a real fleet.

\textbf{Timescale and facility filtering.} Grid-side ramp limits apply at the point of
interconnection, after power conversion and facility equipment, and are often assessed over
seconds to minutes. Several stages low-pass filter GPU power on its way there: the NVML sensor
itself (Section~\ref{sec:setup}); PSU bulk capacitance and voltage regulators, which act at
millisecond scales, below our 0.1\,s resolution; an online double-conversion UPS, which passes
load changes through its rectifier unless energy storage is added to smooth them, as proposed
for training loads \cite{choukse2025powerstabilization}; and cooling and other auxiliary loads, which add a slow
component that chunking does not change. We emulate this by differencing over longer $\Delta$
and by low-pass filtering $D(t)$ with time constant $\tau$ (Table~\ref{tab:sensitivity}). The
$\approx$46\,s traces of Table~\ref{tab:reserve} contain too few independent slow events to
resolve timescales beyond 1\,s, so this table uses a second library whose 400-request trials
leave $\approx$111\,s per trace. The reduction shrinks with timescale but remains significant at
the 95\% level up to $\Delta=2$\,s (13.8\%) and, at both levels, up to $\tau=2$\,s (16.8\%); from
5\,s on the lower bounds fall to about zero or below. Minute-scale
limits such as the 10\,MW/min of Section~\ref{sec:problem} need facility-meter data; our claim is
a fast (sub-second to second) ramp benefit, relevant to fast regulation reserves and to sizing the
energy storage that smooths such ramps \cite{choukse2025powerstabilization}.

\section{Conclusion}
\label{sec:conclusion}

This paper generates real GPU traces to concrete grid-operations optimization. We show that
chunked prefill scheduling can regulate GPU power ramp rate without touching peak power, and
that this ramp-reducing effect grows monotonically with system saturation, confirmed
independently along two axes: concurrency (7.0\% to 34.6\% mean-ramp reduction) and
whale-request load (near zero to 39.6\% across a whale-fraction $\times$ whale-size
grid). We translate this mechanism into an actual operations decision, regulation-reserve
procurement, using a model-free bootstrap directly resampling real measured power traces. For
independent servers, an operator could procure about 24\% less fast-ramping reserve under
chunked scheduling at heavy load and 12\% at moderate load, across reliability levels from 95\%
to 99.9\%, a stability the central limit theorem explains. The result carries over to two-GPU tensor-parallel deployments: about 36\% for a dense 14B model, and a smaller 10\% for a sparse mixture-of-experts model, whose light per-token compute leaves less for chunking to smooth. Because the mechanism
strengthens with saturation, the effect is largest when data centers run hottest. The benefit is
a fast-ramp one: it is significant for ramps resolved at up to 2\,s, and synchronized load
changes, which chunking cannot smooth, raise the absolute reserve far more than chunking lowers
it. Validating the extrapolation against facility-meter data from a multi-server deployment,
including UPS and cooling, and at minute timescales, is the main next step.
\bibliographystyle{IEEEtran}
\bibliography{refs}

\appendices
\section{Robustness and Scope Checks}
\label{app:robustness}

The main body's Section~\ref{sec:load-dependence} already establishes load-dependence along
two axes (concurrency, whale fraction/size) on the current dataset. This
appendix reports further checks, calibrated on the parametric (OU) model of Appendix~\ref{app:ou-model} rather than
the main body's model-free bootstrap: a narrower pilot distribution, an open-loop Poisson
arrival process, a genuine Pareto-tailed whale-size distribution, a whale-free boundary
condition, and a larger model under
real tensor parallelism.
Table~\ref{tab:appendix-summary} summarizes single-server ramp reduction for the three checks
sharing the original-dataset and narrow conditions' calibration window, with each row's
concurrency, whale fraction/size, and ramp metric listed alongside that original condition
(Table~\ref{tab:ramp-orig20}'s caption: concurrency 20, 15\% whale frac., 18k--50k chars) so the
setup of every row can be checked directly rather than taken on faith; the fourth (whale-free)
check is qualitatively different and discussed separately in Section~\ref{app:nowhale}. Every
one of these checks' reserve-procurement translations agrees in direction with the main result:
chunk=512 requiring less reserve than mono, without exception. What these checks establish is
that the \emph{sign} of the effect is robust: chunking reduces ramp rate under every traffic
pattern and setup tested. Table~\ref{tab:ramp-orig20} gives the full per-trial ramp-rate and
near-ceiling duty-cycle breakdown for that original condition, referenced throughout the rest of
this appendix; it shares the same concurrency (20) and whale distribution as the main body's own
condition (Table~\ref{tab:ramp_avg}, Section~\ref{sec:ramp}) but is an independently collected,
smaller (3-trial) dataset, not the same trials.

\subsection{Direction Holds at Every Operational Level Being Tested}
\label{app:reserve-caveats}

\textbf{Chunk=512 requires less reserve than mono at every reliability level, in every one of
the conditions this appendix tests, with no exceptions.} Different budget-choice,
correlation, distribution (narrow or wide variance, Pareto-tail or uniform), arrival rate (Poisson or concurrent), sub-saturated, and TP=2/14B are being tested: all use the
same parametric procedure on Table~\ref{tab:reserve-ou}'s calibration
(Appendix~\ref{app:ou-model}), whose Gaussian tail runs measurably lighter
($\approx6\times$ at p99) than real hardware's, so a specific percentage here would carry more
precision than the underlying model supports. The main body's single-server statistics and its
own model-free Table~\ref{tab:reserve} remain the paper's calibrated quantitative evidence.

We also expect this gap to work in the reported direction's \emph{favor}, though we have not
directly verified it: mono's real dynamics are dominated by rare, large, near-discontinuous
power jumps (one long prefill swings the GPU to near-ceiling almost instantly), exactly the
heavy-tailed behavior a Gaussian process underrepresents, while chunk=512's dynamics are, by
the mechanism this paper studies, already smoothed into many small steps which is closer to what the
parametric model can represent faithfully. If the tail defect suppresses mono's simulated
reserve requirement more than chunk's, the appendix's reserve reductions below would be
\emph{conservative}, understating chunk's true benefit rather than overstating it. We flag this
as a plausible reading of the direction, not a measured correction.

Some qualitative patterns are worth noting alongside the headline direction above. \textbf{Arrival correlation
matters}: at even small correlation probabilities ($s{=}0.02$--$0.05$), the \emph{absolute}
required reserve grows sharply (consistent with Appendix~\ref{sec:cf-level}'s threshold effect),
while the chunk-vs-mono \emph{relative} benefit stays comparatively stable. So any reserve
number in this paper, main body included, is best read as an independent-arrival baseline.
\textbf{Traffic pattern matters little to the direction itself}: the narrower pilot
distribution, open-loop Poisson arrivals, and genuine Pareto-tailed whale sizes (below) all show
chunk=512 requiring less reserve than mono at every reliability level, consistent across
every condition tested, even where we do not quantify by how much.

\begin{table*}[htbp]
\caption{Per-trial ramp rate and near-ceiling duty cycle, by scheduling budget (original dataset,
historical condition referenced throughout the rest of this appendix: widened request-size
distribution, closed-loop, concurrency 20, 15\% whale frac., 18k--50k char / $\sim$5.5k--15.5k
tok.\ whale size, uniform -- an independently collected, smaller 3-trial dataset sharing the
same nominal setup as the main body's own condition, Table~\ref{tab:ramp_avg})}
\label{tab:ramp-orig20}
\centering
\small
\begin{tabular}{@{}lcccccc@{}}
\toprule
Trial & mono ramp (W/s) & 2048 ramp & 512 ramp & mono $\ge$thresh & 2048 & 512 \\
\midrule
1 & 46.0 & 43.8 & 27.8 & 39.7\% & 47.3\% & 59.8\% \\
2 & 46.5 & 43.4 & 34.3 & 40.2\% & 41.0\% & 53.1\% \\
3 & 44.6 & 41.2 & 30.2 & 40.1\% & 46.5\% & 56.6\% \\
\bottomrule
\end{tabular}
\end{table*}

\begin{table*}[htbp]
\caption{Single-server ramp reduction across conditions (mean $\pm$ std, 3 trials each), with
setup parameters listed so each row is directly checkable against the original condition
(Table~\ref{tab:ramp-orig20}: concurrency 20, closed-loop, 15\% whale frac., 18k--50k chars,
uniform, whole-trace ramp metric). Reserve-procurement translations for these conditions are positive
(chunk requires less than mono) at every reliability level tested but are not quantified here
(Appendix~\ref{app:reserve-caveats}).}
\label{tab:appendix-summary}
\centering
\small
\begin{tabular}{@{}lcccc@{}}
\toprule
Condition & Conc.\ / arrival & Whale frac.\ / size & Ramp metric & Reduction \\
\midrule
Widened & 20, closed & 15\% / 18k--50k chars, uniform & whole-trace & 32.7\% $\pm$ 5.4 \\
Narrow & 20, closed & 15\% / $\sim$44k--50k chars, uniform & whole-trace & 34.7\% $\pm$ 5.5 \\
Widened, Poisson & open, rate-matched to conc.\ 20 & 15\% / 18k--50k chars, uniform & whale-window & 42.8\% $\pm$ 14.6 \\
Pareto-tail & 20, closed & 15\% / Pareto $\alpha{=}3.0$, mean $\sim$26k chars & whale-window & 22.8\% $\pm$ 5.9 \\
\bottomrule
\end{tabular}
\\[4pt]
\footnotesize ``whole-trace'' = mean $|dP/dt|$ over the entire trial; ``whale-window'' =
Section~\ref{sec:ramp}'s mechanism-localized measure within whale prefill windows only -- the
two bases are not directly comparable in absolute terms, only in sign and rough size.
\end{table*}

\subsection{Narrower Request-Size Distribution}
\label{app:narrow}

In this experiment, we conduct a pilot parameterization: whale
requests drawn from a near-constant $\sim$44k--50k character range and a short-request population with narrower length variability.
Replicated 3$\times$ with the same paired mono/2048/512 design, whole-trace mean ramp rates are
38.2/35.4/22.2, 41.4/38.6/29.6, and 44.3/43.6/29.4\,W/s (mono/2048/512) across the three
trials, a $34.7\%\pm5.5$ single-server ramp reduction (mono$\to$512, per-trial
41.9/28.5/33.6\%). Under the
parametric model (Appendix~\ref{app:ou-model}), chunk=512 required less reserve than mono
at every reliability level, consistent in direction with the main result. The direction
and rough magnitude of the single-server effect are unchanged by widening the distribution; the main body
uses the wider, more realistic parameterization specifically because it is the more
conservative (smaller-effect) and more representative choice, not because the narrower pilot
disagreed with it.

\subsection{Open-Loop Poisson Arrivals}
\label{app:poisson}

\begin{figure*}[t]
\centering
\includegraphics[width=0.95\textwidth]{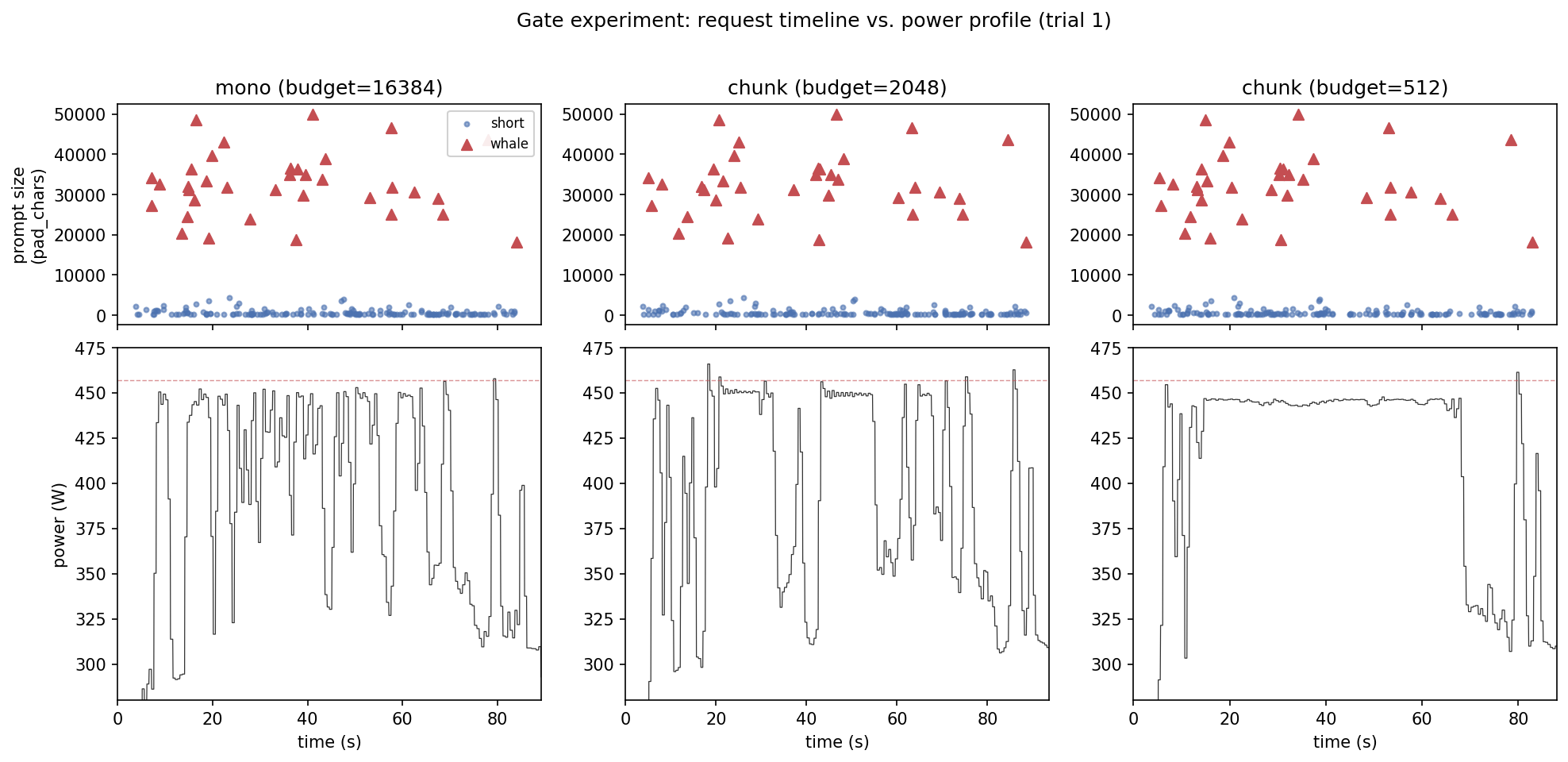}
\caption{Appendix check: widened distribution, open-loop Poisson arrivals. Chunk=512 shows an
especially long, sustained near-ceiling plateau with very few deep excursions -- the
ramp-rate/duty-cycle effect survives a fundamentally different arrival discipline.}
\label{fig:appendix-poisson}
\end{figure*}

In this experiment, we feed the server with the same widened request size distribution but replayed under an open-loop Poisson arrival
process with rate matched to the closed-loop
condition's average throughput. Figure~\ref{fig:appendix-poisson}
shows the timeline/power view: chunk=512 sustains an especially long, near-ceiling plateau with
very few deep excursions. Replicated 3$\times$: single-server ramp reduction $42.8\%\pm14.6$
(per-trial 62.0/39.8/26.7\%); under the parametric model (Appendix~\ref{app:ou-model}),
chunk=512 also required less reserve than mono at every reliability level (not quantified here,
Appendix~\ref{app:reserve-caveats}). The effect survives a fundamentally
different arrival discipline.

\subsection{Pareto-Tailed Whale-Size Distribution}
\label{app:pareto}

\begin{figure*}[t]
\centering
\includegraphics[width=0.95\textwidth]{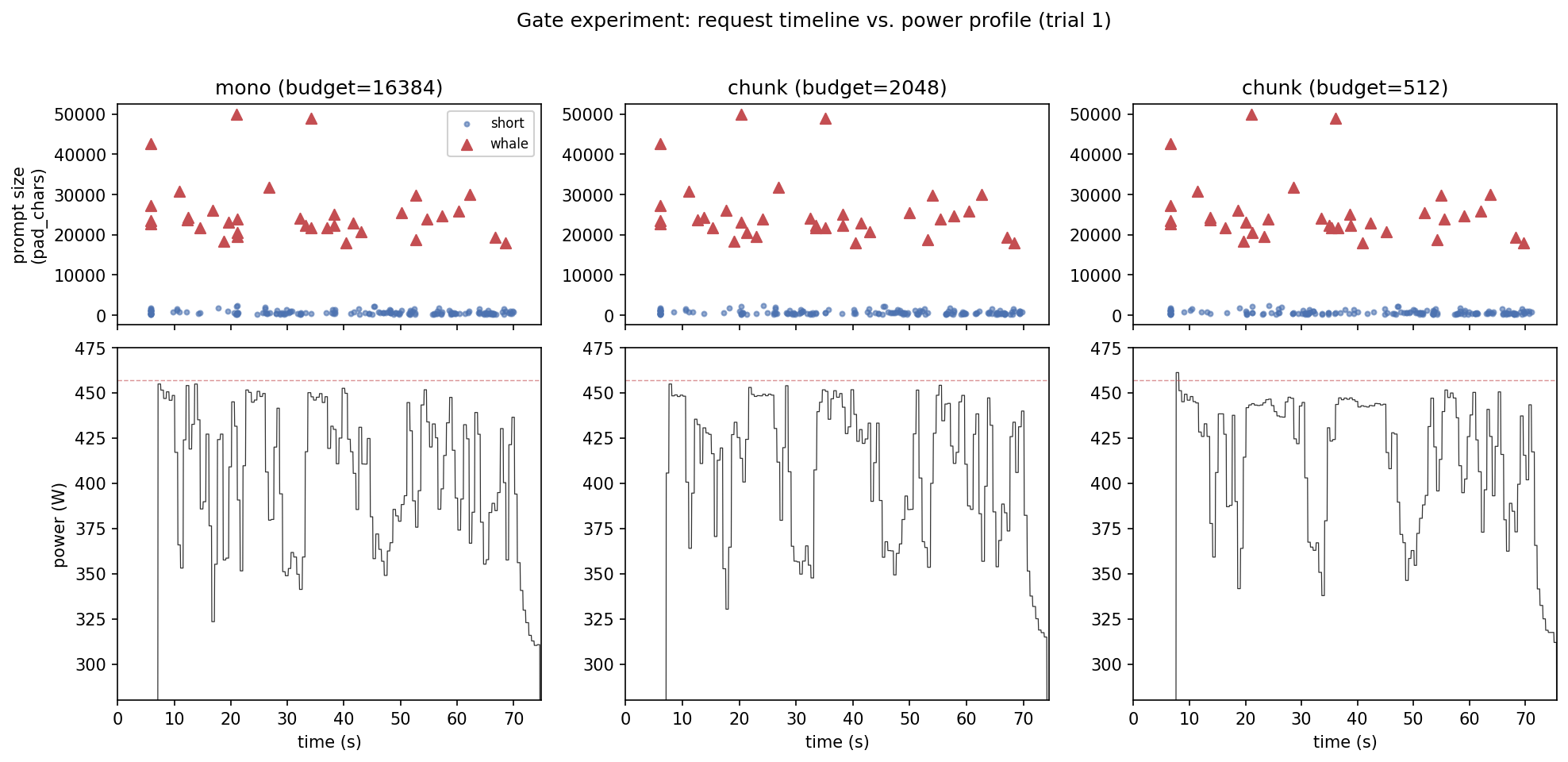}
\caption{Appendix check: genuine Pareto-tailed whale-size distribution ($\alpha{=}3.0$)
instead of uniform. Whale sizes visibly spread across the 18k--50k character range rather than
clustering near the ceiling.}
\label{fig:appendix-pareto}
\end{figure*}

In reality, incoming requests' length likely follow long tail distributions. Here, we set up the traffic such that whale sizes are drawn from a genuine Pareto tail ($\alpha{=}3.0$, $x_m{=}18{,}000$ characters,
capped at 50,000) spliced onto the ordinary lognormal short-prompt body, the standard technique for heavy-tailed
request/flow/file-size traffic (an ordinary body below the crossover, a genuine power-law tail
above it; a single Pareto spanning the full short-to-whale size range would require
$\alpha{<}1$, an infinite-mean, numerically pathological regime). Figure~\ref{fig:appendix-pareto}
shows the timeline/power view, replicated 3$\times$ with single-server ramp reduction
$22.8\%\pm5.9$. Under the
parametric model (Appendix~\ref{app:ou-model}), chunk=512 required less reserve than mono
at every reliability level (Appendix~\ref{app:reserve-caveats}). This
single-server reduction is
meaningfully smaller and noisier than either uniform-whale condition. The Pareto tail's mean
whale size ($\approx$26,000 characters) is smaller than the widened uniform distribution's mean
($\approx$34,000 characters), consistent with an effect whose magnitude scales with how much prefill work there is to spread out rather than merely whether a long-tailed population is
present at all.

\subsection{Whale-Free Boundary Condition}
\label{app:nowhale}

This appendix's checks are measured against the bimodal,
whale-driven mixture (Table~\ref{tab:ramp-orig20}: 85\% short-prompt population, 15\% whale
population, closed-loop concurrency 20). This single-trial check asks whether that benefit
requires the specific short/whale contrast, or would appear under any sufficiently long,
elevated-duration prompt population on its own: we remove the whale population entirely (whale
fraction 0) and replace the whole request stream with one tightly-clustered, individually-long
population, holding concurrency, request count, and every other harness setting identical to
that original condition.

\textbf{Setup.} A single lognormal population, mean 12,000 characters ($CV^2{=}0.08$), truncated
to [7,000, 18,000] characters -- individually long, right up to but never crossing that
condition's 18,000-character whale floor, with no short population mixed in and no bimodal
rare/common contrast. Realized mean prompt size 11,344 characters. Closed-loop concurrency 20, 200
requests/trial, single trial (a boundary/mechanism probe, not a replicated headline number).
Realized mean TTFT climbs to a genuinely multi-second regime (mono mean 0.9\,s, p95 2.1\,s, max
4.5\,s), confirming these prompts are individually long enough to matter.

Table~\ref{tab:nowhale-summary} places this whale-only population's ramp rate and near-ceiling
duty cycle side by side with the original dataset's short+whale mixture
(Table~\ref{tab:ramp-orig20}), both at concurrency 20, so the two setups can be compared
directly rather than only through prose.

\begin{table}[htbp]
\caption{Ramp rate and near-ceiling duty cycle: original dataset's short+whale mixture
(Table~\ref{tab:ramp-orig20}, concurrency 20, mean of 3 trials) vs.\ this appendix's whale-only,
no-short population (concurrency 20, single trial)}
\label{tab:nowhale-summary}
\centering
\footnotesize
\begin{tabular}{@{}lcccc@{}}
\toprule
 & \multicolumn{2}{c}{Short+whale (orig.)} & \multicolumn{2}{c}{Whale-only} \\
Budget & Ramp (W/s) & Duty cycle & Ramp (W/s) & Duty cycle \\
\midrule
mono (16384) & 45.7 & 40.0\% & 36.7 & 58.1\% \\
chunk=2048   & 42.8 & 44.9\% & 36.3 & 65.9\% \\
chunk=512    & 30.8 & 56.5\% & 17.3 & 80.9\% \\
\bottomrule
\end{tabular}
\end{table}

Chunk=2048 vs.\ mono: essentially no change ($-1\%$, $36.7\to36.3$\,W/s) -- the same budget that
drives the main body's primary studied mechanism does nothing here. Chunk=512 vs.\ mono, however, shows a real $53\%$ ramp
reduction ($36.7\to17.3$\,W/s), with duty cycle rising and episode count falling (80.9\% vs.\
58.1\%; 14 vs.\ 32 episodes). This is the \emph{same} direction as the original dataset's own
short+whale mixture (Table~\ref{tab:nowhale-summary}), so duty-cycle direction alone does not distinguish the
two mechanisms. The real difference is \emph{when} the effect turns on. In the short+whale
mixture, even the moderate budget (2048) already reduces ramp, because chunking interleaves a
whale's chunks into decode-only gaps that would otherwise sit idle between bursts. Here there are
no such gaps to fill: every request is already long, so the batch is continuously busy even
under mono, and chunk=2048 changes nothing. Only the aggressive budget (512), which forces
$\sim$7 rounds per request, pins the GPU into a continuous high-power steady state through sheer
round-count at concurrency 20, not through smoothing a spike into an otherwise-idle background.
The resulting ramp-rate number looks similar to the whale-driven case, but the mechanism --
saturation of an already-busy system, rather than smoothing of a rare spike into idle gaps -- is
different.

\textbf{Interpretation.} Burstiness (a rare long prefill against a much shorter background) is
necessary for the spike-smoothing mechanism the main body studies at budget=2048: removing the
whale population kills the effect for that budget entirely, even when every request is
individually lengthened to just under the whale floor. Sufficiently fine-grained chunking
(512) can still reduce ramp in a whale-free, uniformly-long population, but through GPU
saturation, not smoothing, which is a distinct mechanism with a different grid implication in continuous near-ceiling draw rather than intermittent draw.

\subsection{Generalization to a Larger Model, Tensor Parallelism (TP=2), and a Sparse MoE Architecture}
\label{app:tp2-14b}

Section~\ref{sec:ramp}'s single-GPU, 7B-model condition raises an obvious scope question: does
the reserve-procurement benefit require a small model with no cross-GPU communication, or does
it survive at larger scale? We re-run the same widened traffic pattern, at a sub-saturated
calibration (concurrency 4, chosen so whale windows
again sum to $\approx$1$\times$ the trial span), on Qwen2.5-Coder-14B-Instruct split via
tensor parallelism (TP=2) across two RTX 4090s connected by PCIe only (no NVLink), a
host-staged NCCL configuration that, if anything, biases against seeing a clean effect, not in
its favor.
This check also uses the parametric model for data center scale simulation.

\textbf{Result: the reserve-procurement benefit's direction generalizes to this configuration
too}, though on a narrower evidence base (3 trials, one traffic pattern, one budget pair,
parametric model) than the main body, worth treating as an encouraging early signal rather than
a fully powered replication. The underlying single-server mechanism replicates quantitatively as
well: whole-trace mean ramp rate falls from 36.6/35.9/33.4\,W/s (mono) to 27.0/26.6/26.1\,W/s
(chunk=512) across the three trials, a $24.6\%\pm2.0$ single-server ramp reduction (per-trial
26.1/25.8/21.8\%), though smaller than the saturated single-GPU condition's $32.7\%\pm5.4$
(Table~\ref{tab:appendix-summary}) but larger than the sub-saturated single-GPU condition's
$4.5\%\pm1.4$, consistent with this configuration's own sub-saturated (concurrency 4) load level.
Under the parametric model, chunk=512 also required less reserve than mono at every reliability
level for this configuration (not quantified here, Appendix~\ref{app:reserve-caveats}), providing
evidence that a larger model under real tensor parallelism retains the same direction of
benefit as the single-GPU case.

\textbf{A second, complementary mechanism: tensor-parallel communication cost also caps peak
power.} Unlike the single-GPU case, chunking here brings an additional grid benefit on top of
the ramp-rate reduction: peak power itself drops with chunk budget. Table~\ref{tab:tp2-mechanism} places the pooled peak-power change and whale-burst stretch ratio side by side across all
four configurations: chunk=512's peak is lower than mono's both in the per-arm summary and
pooled across all 92 whale events in the sub-saturated 3-trial dataset (mono's mean per-whale
peak $688.8\pm18.1$\,W vs.\ chunk=512's $651.1\pm13.0$\,W) which is a tighter spread, not just a lower
mean, consistent with a genuine cap rather than noise. This complements, rather than contradicts,
the single-GPU finding (Section~\ref{sec:ramp}) that peak power is architecturally insensitive
to budget: tensor parallelism introduces a second lever the single-GPU setting simply does not
have, entirely independent of any model-based reserve calculation.

\begin{table}[htbp]
\caption{Mechanism comparison: single-server ramp reduction (mono $\to$ chunk=512, mean $\pm$
std across 3 trials), pooled peak-power change, and whale-burst stretch ratio, across four
configurations; all rows share the same 15\% whale frac., 18k--50k char whale size, uniform
distribution, closed-loop arrivals}
\label{tab:tp2-mechanism}
\centering
\footnotesize
{%
\begin{tabular}{@{}lcccc@{}}
\toprule
Configuration & Conc. & Ramp reduction & Peak power $\Delta$ & Stretch ratio \\
\midrule
7B, saturated (orig.) & 20 & 32.7\%$\pm$5.4 & $\le$2.3\%    & 2.25$\times$ \\
7B, sub-saturated         & 6 & 4.5\%$\pm$1.4 & $+0.3\%$      & 1.20$\times$ \\
14B/TP2, sub-saturated    & 4 & 24.6\%$\pm$2.0 & $-5.5\%$      & 1.33$\times$ \\
MoE (2.7B active)/TP2     & 20 & 19.8\%$\pm$12.8 & $-1.3\%$    & 1.56$\times$ \\
\bottomrule
\end{tabular}%
}
\end{table}

\textbf{A third generalization: sparse Mixture-of-Experts.} All results above use dense models.
We additionally test Qwen1.5-MoE-A2.7B-Chat (14.3B total parameters, 2.7B activated per token)
under the same TP=2 configuration and widened traffic pattern. At the same sub-saturated
concurrency=4 used for the dense 14B check, chunking's ramp-reduction benefit was much smaller
for this architecture ($2.6\%\pm1.3$ vs.\ the dense model's $24.6\%\pm2.0$ at the same
concurrency), consistent with a lower-activation model simply having less sustained per-token
compute for the interleaving mechanism to act on at light load. Re-running at a heavier load
(concurrency 20, Table~\ref{tab:tp2-mechanism}'s fourth row) recovered most of this gap
($19.8\%\pm12.8$), supporting a load-dependence explanation over an architecture-specific
ceiling -- though this specific comparison mixes both concurrency and architecture at once (we
do not have a dense-model TP=2 data point at concurrency 20 to isolate the two), and the
per-trial spread at concurrency 20 is wide (3 trials: 35.8/19.3/4.4\%). We report this as a
preliminary but encouraging signal that the paper's core mechanism generalizes to sparse
architectures given sufficient load, not a fully resolved comparison.

\textbf{Update in version 2.} Section~\ref{sec:reserve-tp2} now repeats the full model-free
reserve pipeline for both TP=2 models at concurrency 20 with 100 trials per policy, which
supersedes the 3-trial, parametric-model evidence of this appendix for the reserve question and
provides the dense-model concurrency-20 data point that was missing above. It also corrects the
MoE row of Table~\ref{tab:tp2-mechanism}: over 100 trials per policy the MoE concurrency-20
mean-ramp reduction is 4.4\%$\pm$9.7, not 19.8\%$\pm$12.8, so the 3-trial value was at the high
end of the trial-to-trial spread. The MoE benefit at concurrency 20 is only marginally above the
2.6\% at concurrency 4 and remains well below the dense models'. The TP=2 peak-power cap is also smaller at heavy load:
at concurrency 20 the mean per-trial peak falls 1.3\% (14B) and 2.9\% (MoE), against 5.5\% for
14B at concurrency 4.

\textbf{Scope.} Replicated across 3 trials, this single traffic pattern, single budget pair
(mono vs.\ 512), and small set of model/parallelism configurations is enough to confirm that the
reserve-procurement benefit generalizes well beyond the single-GPU, 7B-specific setup studied in
the main body. A natural next step, which we leave to future work, is to fully map the
TP-specific mechanism's dependence on chunk budget, concurrency, TP degree, and a matched
concurrency sweep for the MoE architecture. The dense-model comparison
here is deliberately sub-saturated-to-sub-saturated so as to isolate the effect of model size and parallelism from the
effect of load level studied above; a fleet-scale reserve number under the main body's
saturated methodology is now given in Section~\ref{sec:reserve-tp2}.

\subsection{Latency-Bundling Detail}
\label{app:latency}

\begin{figure}[t]
\centering
\includegraphics[width=\columnwidth]{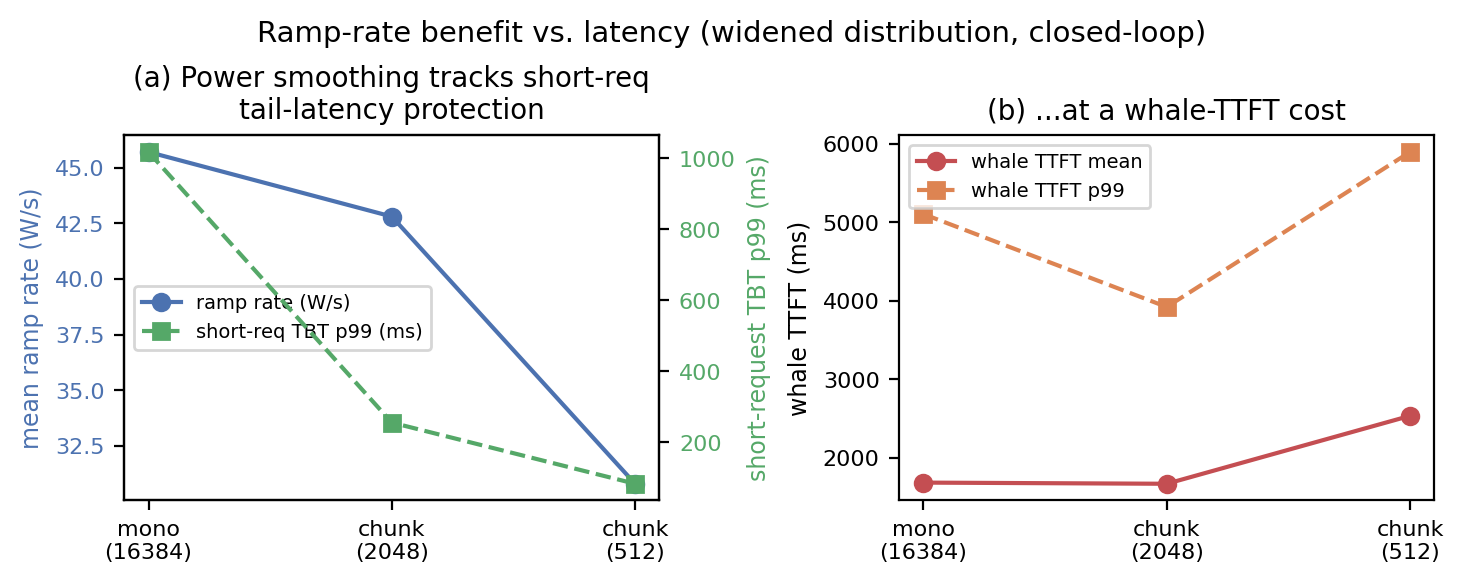}
\caption{The ramp-rate/power-shape benefit is bundled with the already-known short-request
tail-latency protection that motivates chunked prefill (panel a), at a whale-TTFT cost that
is real but non-monotonic in budget (panel b).}
\label{fig:ramp-latency}
\end{figure}

Section~\ref{sec:ramp} notes that chunked prefill's ramp-rate benefit is bundled with the
latency benefit that motivated its deployment in the first place. Figure~\ref{fig:ramp-latency}
gives the full picture, computed from the same 3-trial, original-dataset condition as
Table~\ref{tab:ramp-orig20}:
short-request tail inter-token latency (TBT stands for time-between-tokens, the delay between
consecutive output tokens during generation) falls in lockstep with ramp rate as
chunking gets more aggressive (panel a). As we can see, the power and latency benefits move together, not
independently. The cost is paid by the whale request's own TTFT (time-to-first-token, the
latency until generation begins) (panel b), and non-monotonically: budget=2048 is actually the
best point for whale TTFT, and only the most aggressive setting (512) shows a clear cost.

\section{Parametric Coincidence-Factor Model (Secondary)}
\label{app:ou-model}

Section~\ref{sec:cf-reserve} states why the main body's headline reserve-procurement number
uses a model-free bootstrap rather than a parametric fleet model. This appendix documents that
parametric model in full, which earns its keep because
it yields a closed-form scaling law in Eq.~\eqref{eq:cframp} explaining \emph{why} fleet
averaging attenuates ramp-coincidence risk, a qualitative insight the model-free bootstrap does
not provide. Section~\ref{app:reserve-caveats} discusses what that means for reading them.

\subsection{Coincidence-Factor Grid-Impact Model}
\label{sec:cf}

To translate the single-GPU measurement into a fleet-scale, grid-relevant statement, we build a
Monte Carlo model of $N$ servers and ask how their individually-measured power behavior
aggregates, which is the same ``coincidence factor'' / ``diversity factor'' logic utilities already
used to size distribution equipment for a population of loads rarely all active at once. Each
server is modeled as an alternating-renewal ON/OFF process, where OFF is the decode-only baseline and ON
is the long-prompt prefill, with arrival rate and mean burst duration calibrated \emph{separately
for each scheduling policy from its own trace}, not shared. For example, mono's busy window has duty cycle
$p{=}0.561$ and mean duration $1.63$\,s, while chunk=512's runs measurably longer, $p{=}0.696$
and $3.68$\,s because chunking spreads the same prefill work across more decode-interleaved rounds.
Therefore a whale request occupies the server for longer even though each round is individually
cheaper.

Within each regime, power is modeled as an Ornstein-Uhlenbeck (OU) process, mean-reverting to
a regime-specific level ($\mu_b$ off, $\mu_{max}$ on) with continuous Gaussian noise. The
mean-reversion rate is fit so an isolated server's simulated mean $|dP/dt|$ matches the measured
rate, and Table~\ref{tab:ou-validation} checks
this fit directly against the real single-server ramp-rate distribution: the mean matches by
construction, but real hardware produces occasional larger ramp
events than Gaussian noise reproduces. Comparing the same real single-server trace's own 99th percentile ramp against the 99th percentile of an equal-length OU-simulated trace, calibrated
to match only the mean as above, gives a real/simulated ratio of $\approx6\times$. The OU
process's Gaussian tail decays far faster than the heavier tail real hardware transients
actually show, and this gap is why Section~\ref{sec:cf-reserve} uses a model-free bootstrap for the headline number.

In coincidence factor models, a synchronization parameter
$s\in[0,1]$ sets cross-server correlation ($s{=}0$: independent; $s{=}1$: fully synchronized);
\textbf{every result below uses $s{=}0$ unless a specific $s$-sweep is stated}. Full mechanics and derivations are in Appendix~\ref{app:derivation}.

\begin{table}[htbp]
\caption{Single-server mean ramp rate, real measured trace vs.\ the OU model fit (original
dataset, closed-loop conc.\ 20, 15\% whale frac., 18k--50k char whale size). The calibration
target, matched in both configurations.}
\label{tab:ou-validation}
\centering
\small
\begin{tabular}{@{}lcc@{}}
\toprule
 & real (W/s) & simulated, OU (W/s) \\
\midrule
mono & 46.9 & 45.8 \\
chunk=512 & 27.6 & 30.6 \\
\bottomrule
\end{tabular}
\end{table}

\subsection{Ramp Rate and Level Coincidence Factors}
\label{sec:cf-level}

\begin{figure}[htbp]
\centering
\includegraphics[width=\columnwidth]{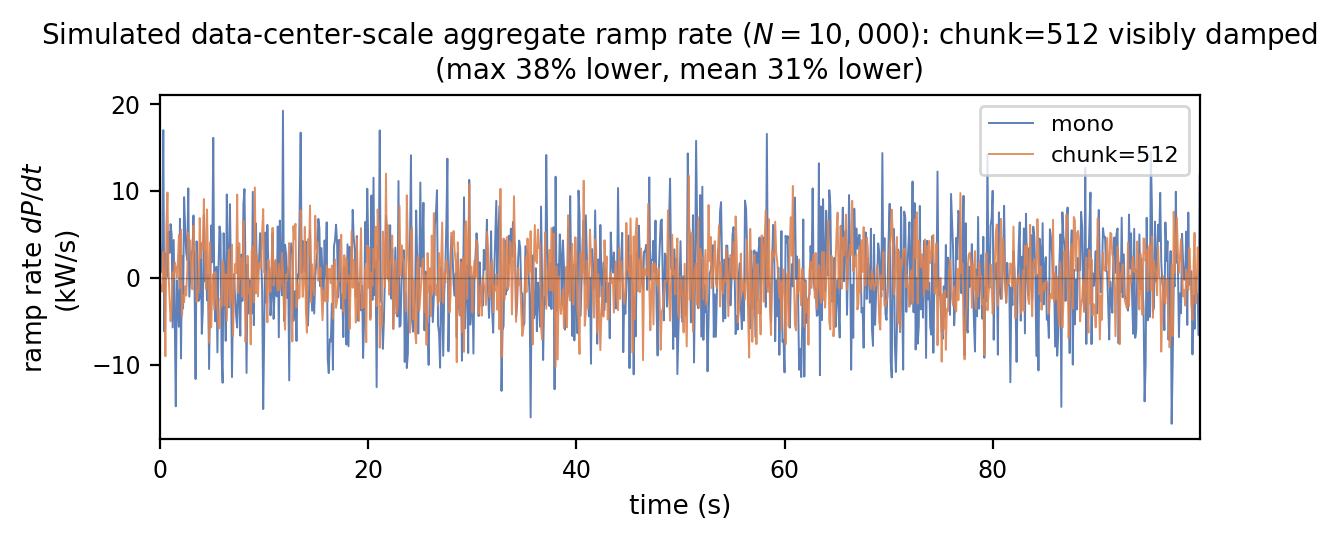}
\caption{A single representative realization of the simulated aggregate ramp rate $dP/dt$ at
$N{=}10{,}000$ servers ($s{=}0$): chunk=512's oscillation is visibly damped throughout (38\%
lower max, 31\% lower mean $\lvert dP/dt\rvert$ in this realization, consistent with
Table~\ref{tab:reserve-ou}'s $\approx$31--32\%) -- this is the quantity the paper's
reserve-procurement argument rests on}
\label{fig:datacenter-profile}
\end{figure}

Figure~\ref{fig:datacenter-profile} shows a representative realization of the simulated
aggregate trace at $N{=}10{,}000$, the fleet size used throughout this subsection and the
reserve-procurement calculation of Section~\ref{sec:cf-reserve-ou} below.

\begin{figure}[htbp]
\centering
\includegraphics[width=\columnwidth]{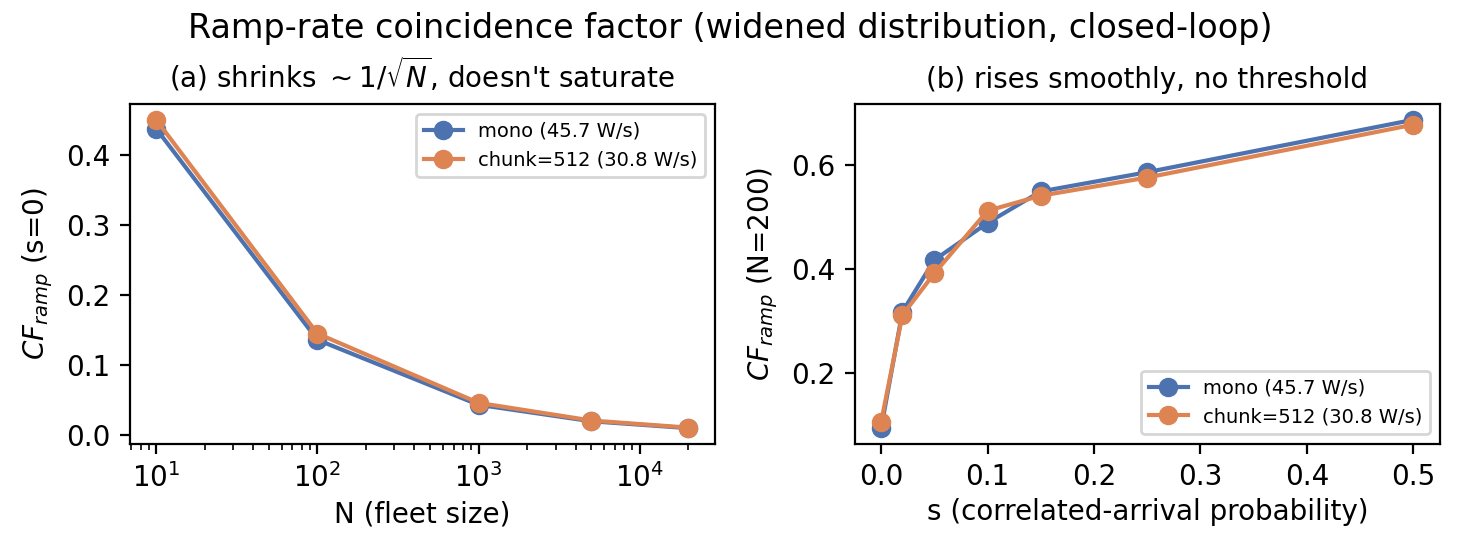}
\caption{Ramp-rate coincidence factor, normalized by each configuration's own single-server
rate. (a) Both curves shrink as $\sim 1/\sqrt{N}$ rather than saturating -- the ramp-coincidence
risk averages out at fleet scale, unlike the level metric's floor in Fig.~\ref{fig:cf-level}.
Chunk=512's \emph{absolute} realized aggregate ramp remains $\approx$31\% lower throughout,
which is the practically relevant quantity. (b) Both configurations rise smoothly with $s$,
unlike the level metric's threshold jump in Fig.~\ref{fig:cf-level}(b).}
\label{fig:cf-ramp}
\end{figure}

\textbf{Ramp rate} scales as $\sqrt N$.
\label{sec:cf-ramp}
Does the single-GPU ramp-rate benefit survive to fleet scale under this parametric model? We
define the ramp-rate analogue
of the level metric, $CF_{ramp}(N) = E[\max_t |D'(t)|]/(N\cdot R)$,  where $R$ is single-server ramp rate. Full derivation is in
Appendix~\ref{app:derivation} via a
random-telegraph-process argument and the functional central limit theorem. Unlike the level metric, it shrinks as the fleet grows:
\begin{equation}
CF_{ramp}(N) \;\sim\; \frac{\sigma}{R}\cdot\frac{M}{\sqrt N},
\label{eq:cframp}
\end{equation}
where $\sigma$ is the single-server ramp-rate standard deviation and $M$ is a fixed constant
independent of $N$, validated by a log-log regression of the simulated $CF_{ramp}(N)$: a fitted
slope of $\approx0.50$ for both mono and chunk=512 (confirming the $1/\sqrt N$ exponent to
within $1.6\%$ residual) and a fitted constant $M\approx3.3$--$3.5$ that agrees closely between
the two configurations. Figure~\ref{fig:cf-ramp} plots both curves. What this scaling law does
and does not imply for the absolute, practically-relevant chunking benefit is discussed in
Appendix~\ref{app:derivation}.

Alongside this ramp-rate generalization, the same $N$-server model surfaces a second,
independent finding about aggregate \emph{peak} demand: unlike ramp rate, the level coincidence
factor converges to a nonzero floor rather than vanishing as the fleet grows, and correlated
arrivals push it toward that floor's ceiling as a threshold rather than a graded effect. Full
derivation, Monte Carlo validation, and the adversarial-risk discussion are in
Appendix~\ref{app:derivation}.

\subsection{Ramping-Reserve Procurement Under the Parametric Model}
\label{sec:cf-reserve-ou}

We solve Eq.~\eqref{eq:reserve-opt} under this parametric model as a point of comparison for
the robustness checks of Appendix~\ref{app:robustness}, simulating $N{=}10{,}000$ servers at $s{=}0$ using
the per-policy-timing, continuous-noise model above, calibrating mono and chunk=512
\emph{separately from each of their own real trials} and collecting the realized peak ramp from 100 independent
100-second windows per trial, then reporting the mean $\pm$ std across different trial-pairs.
Empirical quantiles give $R^*$ at the 95\%/99\% levels; a Gumbel tail fit, validated against
those two empirical quantiles, extrapolates the 99.9\% level (this extrapolation should be read
as directional because real hardware's ramp tail is heavier than the model's, per
Table~\ref{tab:ou-validation}).

\begin{table}[htbp]
\caption{Ramping-reserve capacity required at $N{=}10{,}000$ under the parametric (OU)
model, on the original (pre-diversity-fix) dataset (closed-loop conc.\ 20, 15\% whale frac.,
18k--50k char whale size, same condition as Table~\ref{tab:ramp-orig20}), by reliability level, mono
vs.\ chunk=512 (mean $\pm$ std across 3 trials, each independently calibrated)}
\label{tab:reserve-ou}
\centering
\small
\begin{tabular}{@{}lccc@{}}
\toprule
Reliability & mono & chunk=512 & Reduction \\
 & (MW/min) & (MW/min) & \\
\midrule
95\%   & 1.388 $\pm$ 0.014 & 0.941 $\pm$ 0.017 & 32.2\% $\pm$ 1.9 \\
99\%   & 1.483 $\pm$ 0.017 & 1.022 $\pm$ 0.044 & 31.0\% $\pm$ 3.6 \\
99.9\% & 1.782 $\pm$ 0.048 & 1.216 $\pm$ 0.051 & 31.7\% $\pm$ 3.9 \\
\bottomrule
\end{tabular}
\end{table}

Under this parametric model, chunk=512 requires 31--32\% less ramping-reserve capacity than
mono at every reliability level tested (Table~\ref{tab:reserve-ou}), positive in every trial which is
consistent in direction and rough magnitude with the model-free bootstrap's main-body result
(Table~\ref{tab:reserve}), computed at the same concurrency (20) on an independent dataset and
methodology. Each table's caption
states its own condition in full, for direct comparison. The robustness checks of Appendix~\ref{app:robustness} (budget choice,
correlation, traffic pattern, sub-saturated load, and TP=2/14B) all build on this parametric
model and this table, not on the main body's Table~\ref{tab:reserve}.

\section{Derivation of the Coincidence-Factor Model}
\label{app:derivation}

Appendix~\ref{sec:cf} states the coincidence-factor model's assumptions and two headline
findings in plain terms. This appendix gives the full mechanics and derivations.

\textbf{Simulation mechanics.} For a fleet of $N$ servers, $D(t) = \sum_{i=1}^N P_i(t)$ is the
aggregate power, evaluated over a fixed 100-second reference window (matching the measured
trace's own duration; long enough to contain $\approx$60 renewal cycles for stable Monte Carlo
estimates, short enough that independent servers do not coincide merely because the window was
allowed to grow arbitrarily long, and Appendix~\ref{sec:cf-level} explains why the latter would be
the wrong quantity to compute). Each server's state is the logical OR of a \emph{shared}
component (one realization, occupancy $s\cdot p$, common to all $N$ servers) and an
\emph{idiosyncratic} component (occupancy solved so the union recovers $p$ for every $s$,
independent per server). To capture ramp rate rather than level, each server's instantaneous
ON/OFF step is replaced by a first-order-lag transition,
\begin{equation}
\frac{dP_i}{dt} = \frac{T_i(t) - P_i(t)}{\tau},
\label{eq:lag}
\end{equation}
where $T_i(t)\in\{P_b,P_{max}\}$ is server $i$'s instantaneous ON/OFF target and $\tau =
(P_{max}-P_b)/R$ is calibrated so an isolated server's simulated ramp matches the measured
rate. All ramp-rate results discard a 50-second simulation warm-up (more than 30 relaxation
times) before measuring the peak, avoiding a synchronized-startup transient we identified and
corrected during model validation -- without it, every server begins the window OFF
simultaneously and the whole ensemble's shared relaxation toward equilibrium masquerades as a
coincidence event.

\textbf{Level coincidence factor.} For $N$ servers with uncorrelated arrivals ($s{=}0$), the
level coincidence factor $CF(N) = E[\max_t D(t)]/(N\cdot P_{max})$ converges to a nonzero
floor,
\begin{equation}
CF(N) \;\to\; \frac{P_b}{P_{max}} + \left(1-\frac{P_b}{P_{max}}\right) p,
\label{eq:cf}
\end{equation}
validated by Monte Carlo simulation (predicted 0.923, simulated 0.930 at $N{=}1000$).

\begin{figure}[htbp]
\centering
\includegraphics[width=\columnwidth]{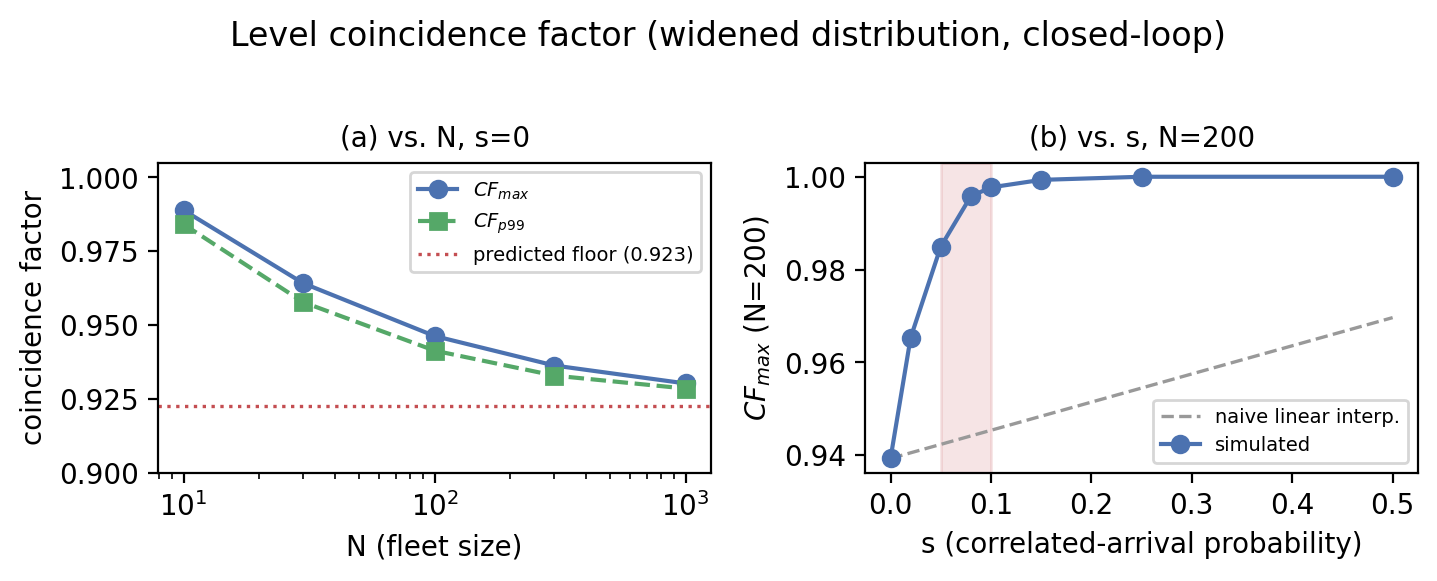}
\caption{Level coincidence factor. (a) Converges to the baseline-corrected floor as $N$
grows, well above the trivial diversity-factor asymptote. (b) Rises far faster than the naive
linear interpolation between the $s{=}0$ floor and $s{=}1$ ceiling -- a threshold, not a
graded, transition.}
\label{fig:cf-level}
\end{figure}

\textbf{Deriving the single-server ramp variance.} Model a single server's ON/OFF indicator
$S(t)$ as a random telegraph process, i.e.\ a two-state Markov chain with stationary occupancy
$p$ and autocovariance
\begin{equation}
\mathrm{Cov}\big(S(t), S(t{+}\tau)\big) = p(1-p)\, e^{-\lambda|\tau|},
\label{eq:telegraph}
\end{equation}
where $\lambda = 1/\text{mean\_on} + 1/\text{mean\_off}$.
Passing the power target $T(t) = P_b + (P_{max}-P_b)S(t)$ through the first-order-lag filter of
Eq.~\eqref{eq:lag} (pole $\kappa=1/\tau$) and applying Campbell's theorem to the resulting
shot-noise-like aggregate gives the single-server ramp-rate variance in closed form:
\begin{equation}
\mathrm{Var}\!\left(\frac{dP}{dt}\right) = \sigma_T^2\,\frac{\lambda\kappa^2}{\lambda+\kappa},
\qquad
\sigma_T^2 = (P_{max}-P_b)^2\,p(1-p).
\label{eq:ramp-var}
\end{equation}
This closed form matches direct Monte Carlo simulation of the filtered process to within
5--15\%.

\textbf{Why the fleet-scale behavior differs from the level metric.} $P_i(t)$ has a nonzero
stationary mean, so by the strong law of large numbers $D(t)/N$ converges pointwise to that
nonzero constant at every $t$ which is exactly the floor in Eq.~\eqref{eq:cf}. In sharp
contrast, $dP_i/dt$ has \emph{zero} stationary mean (a stationary process's derivative averages
to zero), so $D'(t)/N \to 0$: there is no nonzero limit for the level-style argument to
converge to, and all of the fleet-scale ramp behavior is carried by the fluctuation
(central-limit) term around that zero mean, which scales with $N$ differently than a
law-of-large-numbers limit does.

\textbf{The resulting closed form and its exponent.} Because $D'(t)=\sum_i dP_i/dt$ which sums $N$
i.i.d.\ copies of a fixed-statistics process ($s{=}0$), $D'(t)/\sqrt N$ converges, by the
functional central limit theorem, to a fixed, $N$-independent Gaussian process $\sigma Z(t)$,
where $\sigma=\sqrt{\mathrm{Var}(dP/dt)}$ from Eq.~\eqref{eq:ramp-var} and $Z(t)$ is
standardized (unit variance, same correlation shape). Gaussian-process maxima scale
\emph{exactly} linearly under constant rescaling, $\max(cX)=c\max(X)$, for any fixed window
length and \emph{any} correlation shape including the cusped, non-differentiable one
$dP/dt$ actually has. This scale-invariance argument sidesteps that smoothness
requirement entirely and gives the asymptotic law
\begin{equation}
CF_{ramp}(N) \;=\; \frac{E\big[\max_{\text{window}} |D'(t)|\big]}{N\cdot R}
\;\sim\; \frac{\sigma}{R}\cdot\frac{M}{\sqrt N},
\label{eq:cframp-full}
\end{equation}
where $M := E\big[\max_{\text{window}}|Z(t)|\big]$ is a fixed constant -- the expected peak of
the \emph{standardized} process over the 100-second reference window -- independent of $N$.

\subsection{Interpretation: Threshold Risk and the Fate of the Ramp-Rate Benefit at Scale}
\label{app:cf-discussion}

\textbf{Level: correlated arrival is a threshold risk, not a graded one.} Sweeping the
probability $s$ that a given arrival is \emph{shared} (hits every server at once, e.g.\ a viral
prompt or coordinated batch trigger) rather than idiosyncratic, $CF$ does \emph{not} interpolate
smoothly toward the ceiling of 1: any shared event, once it occurs, synchronizes the whole fleet
for its duration, so a sufficiently long window need only contain one. At $N{=}200$, $s{=}0.05$
stays near the independent floor ($CF\approx0.98$) but $s{=}0.10$ already reaches
$CF\approx0.998$, nearly the ceiling (Figure~\ref{fig:cf-level}b). At this calibration,
$s{=}0.10$ corresponds to as little as three or four coordinated, simultaneous long-prompt
submissions per 100-second window. This is structurally the same lesson as cold-load pickup
\cite{alnujaimi2018coldload} where thermostatic loads switch on simultaneously after a feeder
outage.

\textbf{Ramp rate: does the benefit disappear at fleet scale?} Figure~\ref{fig:cf-ramp}(a) shows
both curves shrinking as $\sim1/\sqrt N$ rather than saturating -- so does the ramp-rate benefit
simply disappear at fleet scale? Not quite: because $CF_{ramp}$ is a \emph{ratio} of two
quantities shrinking at the same rate, the \emph{absolute} chunking-vs-mono benefit survives
intact -- chunk=512's realized aggregate ramp stays $\approx$29\% below mono's, stable from
$N{=}100$ through $N{=}20{,}000$ (cf.\ the $\sim$33\% single-GPU figure) -- even though the risk
of many servers' ramps \emph{coinciding} genuinely averages out. Ramp-rate coincidence also rises
smoothly with $s$ rather than jumping at a threshold (Figure~\ref{fig:cf-ramp}b) -- correlated
arrivals still matter for ramp, just without the level metric's abrupt jump.

\textbf{In short:} for this workload and calibration, aggregate peak-demand risk is governed by
the probability of correlated arrivals (a law-of-large-numbers floor that does not average out),
while aggregate ramp-rate risk is governed by ordinary statistical averaging (a
central-limit-theorem fluctuation that does): two structurally different conclusions from the
same model, resolved by asking whether the aggregated quantity has a nonzero or zero stationary
mean.

\end{document}